\documentclass[fleqn,usenatbib]{mnras}

\usepackage{newtxtext,newtxmath}

\usepackage[T1]{fontenc}

\DeclareRobustCommand{\VAN}[3]{#2}
\let\VANthebibliography\thebibliography
\def\thebibliography{\DeclareRobustCommand{\VAN}[3]{##3}\VANthebibliography}

\usepackage{graphicx}	
\usepackage{amsmath}	
\usepackage{comment}

\graphicspath{{Fig/}} 

\title[Tidal decay of planetesimals]{Can tidal evolution lead to close-in planetary bodies around white dwarfs I: Orbital period distribution}

\author[Yuqi Li et al.]{
Yuqi Li,$^{1}$\thanks{E-mail: yl817@cam.ac.uk}
Amy Bonsor,$^{1}$
Oliver Shorttle,$^{1,2}$
and Laura K. Rogers$^{1}$ 
\\
$^{1}$Institute of Astronomy, University of Cambridge, Madingley Road, Cambridge, CB3 0HA, UK\\
$^{2}$Department of Earth Sciences, University of Cambridge, Downing Street, Cambridge, CB2 3EQ, UK}

\date{Accepted XXX. Received YYY; in original form ZZZ}

\pubyear{2024}

\begin{document}
\label{firstpage}
\pagerange{\pageref{firstpage}--\pageref{lastpage}}
\maketitle

\begin{abstract}
The evolution of planetary systems around white dwarfs is crucial to understanding the presence of planetary material in the atmospheres of white dwarfs. These systems uniquely probe exoplanetary compositions. Periodic signals in the photometry of a handful of white dwarfs suggest material blocking the star, potentially from disintegrating planetesimals. Tidal evolution followed by scattering can bring planetesimals onto close-in orbits that would have been within the envelope of the white dwarf progenitor. The orbital period distribution of planetesimals undergoing tidal evolution will peak at short-period (nearly) circularized orbits ($\sim10\,\rm hour$--$1\,\rm day$), with a rising tail towards long-period highly eccentric orbits ($\sim 100\,\rm day$). This prediction is generally consistent with the observed white dwarf transiting systems. In order for the planetesimal on the 4.5\,hour period around WD\,1145+017 to be explained by the tidal evolution of a planetesimal, that planetesimal must have an ultimate tensile strength comparable to that of iron meteorites.

\end{abstract}




\section{Introduction}

White dwarfs that have recently accreted planetary material provide a unique tool to probe the composition of planetary bodies outside the solar system. White dwarfs are the left-over degenerate cores of low to intermediate mass stars. The high surface gravity of white dwarfs leads to rapid gravitational settling of metals (elements heavier than helium). However, observations reveal that $\sim$ 10\%--50\% of the white dwarfs are contaminated with metals, suggesting recent/ongoing accretion of the remaining planetary bodies around the white dwarfs \citep{2003ApJ...596..477Z,2010ApJ...722..725Z,2014A&A...566A..34K,2019MNRAS.487..133W,2023MNRAS.518.3055O,2024MNRAS.527.8687O,2024MNRAS.531L..27M}. 

The mechanism that leads to the accretion of planetary material onto white dwarfs remains unclear. The widely accepted theory is that planetesimals (planetary building blocks) are gravitationally scattered towards the white dwarf under the effect of massive perturbers, for instance, planets \citep{2012A&A...548A.104B,2018MNRAS.476.3939M,2024MNRAS.527.11664}. These scattered planetesimals end up entering the white dwarf atmosphere. The accretion onto white dwarfs can be observed in action, with both dusty material (infrared excess) and gas (circumstellar absorption/emission features) detected \citep{1987Natur.330..138Z,2005ApJ...632L.119B,2005ApJ...632L.115K,2006ApJ...646..474K,2007MNRAS.380L..35G,2007ApJ...663.1285J,2009ApJ...694..805F,2009AJ....137.3191J,2010ApJ...714.1386F,2010ApJ...722.1078M,2012ApJ...754...59D,2015ApJ...806L...5X,2017MNRAS.468..154B,2020ApJ...902..127X,2021ApJ...920..156L,2023ApJ...944...23W} (see \citealp{2016NewAR..71....9F} for a detailed review). 

Further observational evidence for accretion in action comes from the first identification of transits around a polluted white dwarf with an infrared excess, WD\,1145+017 using the Kepler K2 data \citep{2015Natur.526..546V}. Follow-up observations from 2015--2017 re-confirm the dominant period in K2 data, the 4.5\,hr period (`A' period), with variable weaker periodic signals \citep{2015Natur.526..546V,2016ApJ...818L...7G,2016MNRAS.458.3904R,2017ApJ...836...82C,2017MNRAS.465.3267G}.

There is a rich literature that attempts to explain the origin of the transiting features of WD\,1145+017. One popular model motivated by the longer egress time and unexpectedly long transit duration consistent with the light curves of disintegrating planets around main sequence stars \citep{2013ApJ...776L...6K,2014A&A...572A..76V,2014ApJ...784...40R,2015ApJ...812..112S,2016A&A...596A..32V} is a planetesimal with active dust/gas production orbiting the white dwarf with a period of 4.5\,hr \citep{2016MNRAS.458.3904R, 2017MNRAS.465.1008V, 2020ApJ...893..166D, 2020MNRAS.498.4005O}.

The second identified transiting white dwarf system, ZTF\,J0139+5245, possesses a much longer period (107.2\,day) than WD\,1145+017 \citep{2020ApJ...897..171V}. 5 more potential white dwarf transiting systems are identified \citep{2021ApJ...912..125G}, among which ZTF\,J0328-1219 with robust evidence of transits is further studied and two transiting periods 9.937\,hr and 11.2\,hr are found \citep{2021ApJ...917...41V}. Meanwhile, another system with transits, WD\,1054-226 (period 25.02\,hr) is reported \citep{2022MNRAS.511.1647F}. With the increase in publicly available light curves from past and present facilities such as: Kepler/K2 \citep{2010Sci...327..977B,2014PASP..126..398H}, the Transiting Exoplanet Survey Satellite \citep[\textit{TESS},][]{2014SPIE.9143E..20R}, \textit{Gaia} \citep{2016A&A...595A...1G}, and the Zwicky Transient Facility \citep[ZTF,][]{2019PASP..131a8002B}, and future facilities such as: the Large Synoptic Survey Telescope \citep[LSST,][]{2019ApJ...873..111I}, more transit candidates should be identified, with extensive high-speed photometric follow-ups usually required to confirm the transits and obtain the transit periods (e.g., \citealp{2021ApJ...912..125G,2021ApJ...917...41V,2022MNRAS.511.1647F}). 

The transiting systems provide clear observational evidence for planetary material close to the white dwarfs, potentially undergoing active dust/gas production. If the planetary material has not undergone a common envelope event, orbital decay/circularization after gravitational scattering may be required to bring planetary material close to the white dwarfs, with tidal evolution a potential mechanism \citep{2019MNRAS.489.2941V,2020MNRAS.492.6059V,2020MNRAS.498.4005O}.

In the solar system, the tidal effect is responsible for the spin-down of the Earth accompanied with the outward migration of the moon, which help constrain the origin and the evolution history of the Earth-moon system \citep{1966RvGSP...4..411G,KAGAN1997109,1997A&A...318..975N,2015E&PSL.427...74Z,Tyler_2021,2022A&A...665L...1F,2023Icar..38915257D}. Tidal interactions also give rise to the solar system's volcanically active moon, Io \citep{2015ApJS..218...22T,2018AJ....156..207R,2022Icar..37314737K,2024NatAs...8...94D,2024ApJ...961...22S}. Meanwhile, tidal theory has also been applied to bodies outside the solar system, for instance, eccentric migration potentially responsible for the formation of hot (and sometimes inflated) Jupiters \citep{2009ApJ...702.1413M,2010ApJ...713..751I,2010A&A...516A..64L,2021ApJ...920L..16D,2022ApJ...931...10R,2022ApJ...931...11G,2023ApJ...943L..13V}. \citealp{2019MNRAS.486.3831V} investigated the parameter space where tidal evolution brings planets into the Roche limit of white dwarfs. \citealp{2019MNRAS.489.2941V,2020MNRAS.492.6059V} and \citealp{2020MNRAS.498.4005O} incorporated tidal evolution to explain the (potential) presence of close-in planetary bodies around white dwarfs and constrain the tidal-related parameters of the bodies based on their current orbits and the cooling ages of the white dwarfs.

In this paper, following the idea that tidal evolution may lead to close-in planetary bodies around white dwarfs, we examine the scenario that a population of surviving planetesimals scattered onto highly eccentric orbits external to, and close to the Roche limit of the white dwarfs tidally evolve onto various shorter-period orbits. By tracing the orbital parameters of scattered planetesimals undergoing tidal evolution, we:

\begin{itemize}
    \item investigate the parameter dependence of tidal circularization timescale, 
    \item predict the distribution of orbital periods shaped by tidal evolution at a given epoch,
    \item compare the predicted orbital period distribution to periodic signals seen for white dwarfs with transit features
    \item constrain the physical properties of the transiting planetesimals and compare to Solar System asteroids.
\end{itemize}

The paper starts by summarising the model used to follow the tidal evolution of scattered planetesimals (Section \ref{method}) and predict the resultant orbital period distribution. The tidal evolution of an individual planetesimal of given properties is presented in Section \ref{tidal evolution under the CTL model}, highlighting the key factors affecting the tidal circularization timescale (Section \ref{circularization timescale}). By considering the probability of scattering planetesimals to different initial pericentres, the probability distribution of the orbital periods is calculated in Section \ref{period distribution}. Then, we discuss the limitations in the tidal model (Section \ref{tidal model discussion}) and the uncertainties in the population of planetesimals scattered close to the white dwarf (Section \ref{free parameters}), together with the corresponding effects on the orbital period distribution. In Section \ref{period distribution discussion}, we present a synthetic orbital period distribution accounting for a population of planetesimals with a range of properties scattered at different times and compare to current observations. Finally, we discuss the implications of the model in Section \ref{implication} and in Section \ref{conclusion}, we summarize our results.

\section{Methods}\label{method}

Towards the end of the asymptotic giant branch/start of the white dwarf phase, the host star loses a substantial amount of its mass, weakening its gravitational attraction. As a result, the planetary system tends to become less stable against gravitational perturbations \citep{10.1111/j.1365-2966.2011.18524.x,2016RSOS....350571V,2022MNRAS.513.4178O}. Under the growing gravitational effects of perturbers (e.g., planets), the orbital parameters of planetary bodies are altered (via, e.g., scattering), leading to phenomena such as ejection and tidal disruption, with the latter a plausible pathway of polluting the photosphere of the white dwarf \citep{2018MNRAS.476.3939M,10.1093/mnras/stab1667,2023MNRAS.518.4537V,2024MNRAS.527.11664,2024RvMG...90..141V}. In this paper, we will use the term scattering to refer to the delivery of planetesimals close to the white dwarf, but we acknowledge the existence of other mechanisms (e.g., secular chaos \citealp{2022MNRAS.513.4178O}, mean motion resonances \citealp{2023MNRAS.518.4537V}).

The method presented in this work utilises the constant time lag (CTL) model of tidal evolution to predict the orbital evolution of planetesimals scattered close to the white dwarf after leaving the instability zone (which we define as $t=0$). The aim is to predict the orbital period distribution of planetesimals after a certain time of tidal evolution, based on an assumption for the scattering process.

In this section, we:

\begin{itemize}
    \item summarizing the CTL model (Section \ref{equilibrium tide method}), outlining the coupled evolution equations (Section \ref{evolution equations}) and listing the key properties of the model (Section \ref{model properties}),
    \item derive the Roche limit accounting for the ultimate tensile strength of the planetesimal and pseudo-synchronous spin predicted by the constant time lag model, which constrains the orbital parameter space where a planetesimal avoids tidal disruption (Section \ref{Roche limit method}),
    \item summarize the method and list the choice of free parameters (Section \ref{method summary}).
\end{itemize}

\subsection{The constant time lag model}\label{equilibrium tide method}

In this study, we use the CTL model to predict the orbital evolution of planetesimals scattered close to the white dwarf, external to the Roche limit (Section \ref{Roche limit method}), after the planetesimals leave the instability zone (where scattering occurs). Under the CTL model, the tidal force exerted by the white dwarf raises tidal bulges on the planetesimal that lag behind the equipotential surface for a constant time interval $\Delta t$, inducing angular momentum transfer and dissipation of orbital energy  \citep{1981A&A....99..126H,2007A&A...462L...5L,2010A&A...516A..64L,2010ApJ...725.1995M,2011A&A...535A..94B,2011A&A...528A..27H,2012ApJ...751..119B,2012ApJ...757....6H,2022ApJ...931...11G,2022ApJ...931...10R,2023ApJ...948...41L}.  Although more realistic models exist, the unconstrained properties of exoplanetary bodies and the complex coupling between rheological and tidal evolution also add more uncertainties (Section \ref{tidal model discussion}).

\subsubsection{Evolution equations}\label{evolution equations}

\begin{table}
\begin{center}
\begin{tabular}{|c|c| } 
 \hline
 
Subscripts&\\
\hline 

$p$ & planetesimal \\

$*$ & white dwarf\\

$0$ & $t=0$, the point of leaving the scattering zone \\
& orbital evolution dominated by tide\\

$\mathit{eq}$ & equilibrium state under the CTL model\\

\hline

Constants&\\
\hline 

$G$ & gravitational constant\\
$M_{\odot}$& solar mass\\
\hline

Constant parameters&\\
\hline

$M$& mass \\ 

$R$ & radius\\

$\rho$& bulk density\\

$k_2$ & potential love number of degree 2\\

$\Delta t$& constant time lag\\

$K$& $K\equiv 3k_2\Delta t$\\

$I$ & moment of inertia\\

$\sigma_s$ & ultimate tensile strength \\

$C$ & $C\equiv \frac{I}{MR^2}$\\

$T_p$ & $T_p\equiv\frac{K_p(M_p+M_*)M_*R_p^5}{M_p}\approx \frac{K_pM_*^2R_p^5}{M_p}$\\

\hline
Orbital parameters&\\
\hline

$e$& eccentricity\\

$a$& semi-major axis\\

$T$ & orbital period\\

$q$ & pericentre distance\\
& $q=a(1-e)$\\

$Q$ & apocentre distance\\
& $Q=a(1+e)$\\

$\omega$ & spin\\

& $\mathrm{sgn}(\omega)=\mathrm{sgn}(\cos\epsilon)$\\

$\epsilon$ & obliquity\\

$r_{\mathit{Roche}}$ & Roche limit\\

$n$ & mean motion\\
& $n=\sqrt{\frac{G(M_p+M_*)}{a^3}}$\\

$\eta$ & ratio of spin to orbital angular momentum\\
& $\eta\equiv C\frac{M_p+M_*}{M_pM_*}M\frac{R^2}{a^2}(1-e^2)^{-\frac{1}{2}}\frac{\omega}{n}$\\
\hline
Time&\\
\hline
$t$& tidal evolution time \\ 
$t=0$& the time of leaving the scattering zone\\
$\tau_{\mathit{cir}}$ &Tidal circularization timescale\\
\hline
\end{tabular}
\end{center}
\caption{Definitions of the symbols used in this paper.}
\label{definition table}
\end{table}

The coupled tidal evolution equations for a white dwarf-planetesimal (two-body) system, expanded to the lowest order in $\Delta t$ and to the fifth order in $\frac{R}{r}$ are of the form \citep{1981A&A....99..126H,2007A&A...462L...5L,2010A&A...516A..64L,2010ApJ...725.1995M,2011A&A...535A..94B,2011A&A...528A..27H,2012ApJ...751..119B,2012ApJ...757....6H,2022ApJ...931...11G,2022ApJ...931...10R,2023ApJ...948...41L} (with the symbols defined in Table \ref{definition table}):

\begin{equation}\label{tidal e general equation}
\begin{aligned}
&\frac{de}{dt}=\sum_{i=p,*}9K_i n\frac{M_pM_*}{M_i^2}\frac{R_i^5}{a^5}e(1-e^2)^{-\frac{13}{2}}\\&\times\left[\frac{11}{18}(1-e^2)^{\frac{3}{2}}f_4(e)\omega_i\cos \epsilon_i-f_3(e)n\right],
\end{aligned}
\end{equation}

\begin{equation}\label{tidal a general equation}
\begin{aligned}
&\frac{da}{dt}=\sum_{i=p,*}2K_in\frac{M_pM_*}{M_i^2}\frac{R_i^5}{a^4}(1-e^2)^{-\frac{15}{2}}\\&\times\left[(1-e^2)^{\frac{3}{2}}f_2(e)\omega_i\cos \epsilon_i-f_1(e)n\right],
\end{aligned}
\end{equation}

\begin{equation}\label{tidal w equation}
\begin{aligned}
&\frac{d\omega_i}{dt}=K_in^2\frac{(M_pM_*)^2}{(M_p+M_*)M_i^3}\frac{R_i^3}{a^3}\frac{1}{C_i}(1-e^2)^{-6}\\&\times\left[f_2(e)n\cos \epsilon_i-\frac{1}{2}(1+\cos^2 \epsilon_i)(1-e^2)^{\frac{3}{2}}f_5(e)\omega_i\right],
\end{aligned}
\end{equation}

\begin{equation}\label{tidal obliquity equation}
\begin{aligned}
&\frac{d \epsilon_i}{dt}=K_in^2\frac{(M_pM_*)^2}{(M_p+M_*)M_i^3}\frac{R_i^3}{a^3}\frac{\sin \epsilon_i}{\omega_i C_i}(1-e^2)^{-6}\\&\times \left[\frac{1}{2}(\cos \epsilon_i-\eta_i)(1-e^2)^{\frac{3}{2}}f_5(e)\omega_i-f_2(e)n\right],
\end{aligned}   
\end{equation}

\noindent where the subscript $i$ represents the contribution of the tide induced on the object $i$ by the other object to the evolution of the system, and $f_1$ to $f_5$ are of the form:

\begin{equation}\label{eccentricity functions}
\begin{aligned}
&f_1(e)=1+\frac{31}{2}e^2+\frac{255}{8}e^4+\frac{185}{16}e^6+\frac{25}{64}e^8,\\
&f_2(e)=1+\frac{15}{2}e^2+\frac{45}{8}e^4+\frac{5}{16}e^6,\\
&f_3(e)=1+\frac{15}{4}e^2+\frac{15}{8}e^4+\frac{5}{64}e^6,\\
&f_4(e)=1+\frac{3}{2}e^2+\frac{1}{8}e^4\\
&f_5(e)=1+3e^2+\frac{3}{8}e^4.
\end{aligned}
\end{equation}

For a planetesimal-white dwarf system considered in this work, we focus on the tide raised on the planetesimal by the white dwarf (planetesimal tide, $i=p$ terms) because the contributions of white dwarf tide is negligible: the ratio of the contributions of $i=p$ terms to that of $i=*$ terms can be approximated as $\frac{K_p}{K_*}\frac{\rho_*^2R_*}{\rho_p^2R_p}\sim \frac{K_p}{K_*}10^{11}$ with $\log_{10}\frac{K_p}{K_*}\sim 10$ \citep{2010ApJ...713..239W,2019MNRAS.489.2941V,2023ApJ...945L..24B}. 

\subsubsection{Properties of the model}\label{model properties}

\begin{table}
\begin{center}
\begin{tabular}{|c|c| } 
\hline
Equilibrium&\\
\hline

$\omega_{\mathit{eq}}$ & $\frac{2\cos{\epsilon_p}}{1+\cos^2\epsilon_p}\frac{f_2(e)}{(1-e^2)^{\frac{3}{2}}f_5(e)}n$\\
&$\left[G(M_p+M_*)\right]^{\frac{1}{2}}\left(\frac{q_0+Q_0}{2q_0Q_0}\right)^{\frac{3}{2}}\frac{2\cos{\epsilon_p}}{1+\cos^2\epsilon_p}\frac{f_2(e)}{f_5(e)}$\\
& $\mathrm{sgn}\left(\frac{\partial \omega_{\mathit{eq}}}{\partial e}\right)=\mathrm{sgn}(e)$\\

$\epsilon_{\mathit{eq}}$ & $
    \begin{cases}
    0 & 0\leq \epsilon_{\mathit{p,0}}< \frac{\pi}{2}\\
    \pi & \frac{\pi}{2}<\epsilon_{\mathit{p,0}}\leq \pi
    \end{cases} $\\

\hline

Circularization&\\
\hline

$e_{\mathit{cir}}$&0\\

$a_{\mathit{cir}}$&$\frac{2q_0Q_0}{q_0+Q_0}$\\

$\omega_{\mathit{cir}}$ & $n$ \\

$\epsilon_{\mathit{cir}}$ & $ \epsilon_{\mathit{eq}}$\\

\hline

Evolution trends&\\
\hline
$a$& $\mathrm{sgn}(\frac{da}{dt})=-\mathrm{sgn}(e)$\\
& $\mathrm{sgn}(\frac{\partial \dot{a}}{\partial e})=-\mathrm{sgn}(e)$\\
& $\frac{da}{dt}(e\rightarrow 0)=0$\\
$e$& $\mathrm{sgn}(\frac{de}{dt})=-\mathrm{sgn}(e)$\\
& $\mathrm{sgn}(\frac{\partial \dot{e}}{\partial e})=\mathrm{sgn}(e-0.658)$\\
& $\frac{de}{dt}(e\rightarrow 0)=\frac{de}{dt}(e\rightarrow 1)=0$\\
$\omega_{p}$& $\mathrm{sgn}(\frac{d\omega_p}{dt})=\mathrm{sgn}(\omega_{\mathit{eq}}-\omega_p)$\\
$\epsilon_p$& $\mathrm{sgn}(\frac{d\epsilon_p}{dt})=\mathrm{sgn}(\epsilon_{\mathit{eq}}-\epsilon_p)$\\
& $\mathrm{sgn}(\frac{\partial \dot{\epsilon}_p}{\partial \omega_p})=\mathrm{sgn}\left[\frac{2}{\eta_p}\omega_{\mathit{eq}}(\epsilon_p=0)-\omega_p\right]$\\
$q$& $\mathrm{sgn}(\frac{dq}{dt})=\mathrm{sgn}(e)$\\
$Q$ & $\mathrm{sgn}(\frac{dQ}{dt})=-\mathrm{sgn}(e)$\\
\hline

\end{tabular}
\end{center}
\caption{The equilibrium spin, obliquity, and the eccentricity, semi-major axis after tidal circularization under the CTL model, together with the general evolution trends (the signs of the differential equations) of the model. $\mathrm{sgn}(f)$ represents the sign of $f$, which is 1 if $f$ is positive, 0 if $f$ is 0 and -1 if $f$ is negative. See appendices A and B for detailed analysis.}
\label{equilibrium value table}
\end{table}

The key properties of the CTL model \citep{1981A&A....99..126H,2007A&A...462L...5L,2010A&A...516A..64L,2010ApJ...725.1995M,2011A&A...535A..94B,2011A&A...528A..27H,2012ApJ...751..119B,2012ApJ...757....6H,2022ApJ...931...11G,2022ApJ...931...10R,2023ApJ...948...41L} for a white dwarf-planetesimal system are summarized below (see appendices A, B, and see Table \ref{equilibrium value table} for a summary):

\begin{itemize}

    \item The orbital angular momentum of the white dwarf-planetesimal system is conserved during tidal evolution, such that semi-major axis and eccentricity evolution is constrained by the initial pericentre distance ($q_0$) and apocentre distance ($Q_0$) via:
    \begin{equation}\label{a-e relation from constant L}
    \begin{aligned}
    &a(1-e^2)=\frac{2q_0Q_0}{q_0+Q_0}.
    \end{aligned}
    \end{equation}
    
    \item Under the CTL model, tidal evolution starting with identical orbital parameters ($q_0$, $Q_0$) forms a set of equivalent evolution tracks that follow the identical trajectory in $a$--$e$ space and simultaneous in $T_pt$ space, with $T_p$ modulating the tidal evolution rate given by:

    \begin{equation}\label{Tp equation}
     T_p=\frac{K_pM_*^2R_p^5}{M_p}\propto \frac{K_pM_*^2R_p^2}{\rho_p}.
    \end{equation}

    \item In comparison to the timescale of semi-major axis and eccentricity decay, the planetesimal reaches pseudo-synchronization ($\omega_p=\omega_{\mathit{eq}}\propto n$)
    and spin-orbit (mis)alignment ($\epsilon_p=\epsilon_{\mathit{eq}}=0,\pi$) rapidly.
     
   \item The semi-major axis decay rate $|\frac{da}{dt}|$ 
 and orbital period decay rate $|\frac{dT}{dt}|$ increase with eccentricity $e$, and hence declines during tidal evolution.
   
    \item The eccentricity decay rate $|\frac{de}{dt}|$ has its maximum at $e\approx 0.658$.

    \item A planetesimal evolves much faster under tide if it starts at a smaller pericentre distance (for an initially highly eccentric orbit, the analytical circularization timescale satisfies $\tau_{\mathit{cir}}\propto q_0^{7.5}$).

     \item The general trend of tidal evolution of a white dwarf-planetesimal system under the CTL model is the decay in eccentricity, semi-major axis and apocentre distance, the increase in the pericentre distance, accompanied with spin-orbit (mis)alignment and pseudo-synchronization. 
    
    \end{itemize}

\subsection{The Roche limit}\label{Roche limit method}

If the planetesimal is scattered too close to the white dwarf, the differential tidal force may lead to rapid disintegration of the body, in which case tidal evolution is irrelevant. The critical distance where the gravitational acceleration, the maximum acceleration provided by ultimate tensile strength, the centrifugal acceleration and the tidal acceleration add up to 0 is defined as the Roche limit ($r_{\mathit{Roche}}$). Within the Roche limit, the net acceleration leads to fracture of the body, referred to as tidal disruption. Tidal disruption ceases at the point the size of the constituent is reduced to the point where the ultimate tensile strength is sufficient to support the object. In this study, we only consider the planetesimals outside the Roche limit. We apply the model in \citealp{2015MNRAS.450.4233B,2017MNRAS.468.1575B,2022MNRAS.509.2404B}, where the acceleration balance is investigated at the surface (an alternative model considering a fracture plane is discussed in \citealp{1999Icar..142..525D}), further incorporating the pseudo-synchronous spin of the planetesimal undergoing tidal evolution. The balance among the four accelerations for a spherical rigid planetesimal (we will extend our study to non-spherical bodies in Section \ref{roche limit discussion}) can be expressed as:

\begin{equation}\label{balance tidal acceleration}
\begin{aligned}
\\&\frac{2GM_*R_p}{r_{\mathit{Roche}}^3}+\omega_p^2R_p=\frac{GM_p}{R_p^2}+\frac{\sigma_s \pi R_p^2}{M_p},
\end{aligned}
\end{equation}

\noindent where $\sigma_s$ is the ultimate tensile strength. 

The Roche limit can be obtained by solving Eq.\ref{balance tidal acceleration}:

\begin{equation}\label{roche limit general equation}
\begin{aligned}
&r_{\mathit{Roche}}=\left[\frac{2GM_*}{\frac{4}{3}\pi G\rho_p+\frac{3\sigma_s}{4 \rho_pR_p^2}-\omega_p^2}\right]^{\frac{1}{3}}.
\end{aligned}
\end{equation}

Due to the fact that on the timescales of pseudo-synchronization ($\omega_p=\omega_{\mathit{eq}}$) and spin-orbit alignment ($\epsilon_p=0, \pi$), $a$ and $e$ stays nearly constant, the necessary condition of avoiding tidal disruption after scattering of the planetesimal, $q_0> r_{\mathit{Roche,0}}$ (with $q_0$ the initial pericentre distance and $r_{\mathit{Roche,0}}$ the initial Roche limit after scattering), can be obtained by substituting $\omega_p=\omega_{\mathit{eq}}(a_0,e_0,\epsilon_p=0,\pi)$ (see Table \ref{equilibrium value table} for the expression), and solve the following set of equations:

\begin{equation}\label{solve roche limit equation}
\begin{cases}
|\omega_p|=\sqrt{\frac{4\pi G\rho_p}{3}+\frac{3\sigma_s}{4\rho_pR_p^2}-\frac{2GM_*}{r_{\mathit{Roche,0}}^3}}\\|\omega_p|=\frac{f_2(e_0)}{(1-e_0^2)^{\frac{3}{2}}f_5(e_0)}n_0
\\e_0=\frac{Q_0-r_{\mathit{Roche,0}}}{Q_0+r_{\mathit{Roche,0}}}
\end{cases}
,
\end{equation}

Eq.\ref{solve roche limit equation} cannot be solved analytically. However, one can find an approximated solution using the fact that $r_{\mathit{Roche,0}}\ll Q_0$ ($e_0\approx 1-\frac{2r_{\mathit{Roche,0}}}{Q_0}$, $a_0\approx \frac{Q_0}{2}$) usually applies:

\begin{equation}
|\omega_{p}|=|\omega_{\mathit{eq,0}}|\approx \frac{33}{40}\sqrt{2GM_*}{r_{\mathit{Roche,0}}}^{-\frac{3}{2}}, 
\end{equation}

\noindent and the Roche limit can be approximated as:

\begin{equation}\label{roche limit approximated equation}
\begin{aligned}
&r_{\mathit{Roche,0}}\approx \left(\frac{3.36125GM_*}{\frac{4}{3}\pi G\rho_p+\frac{3\sigma_s}{4 \rho_pR_p^2}}\right)^{\frac{1}{3}},
\end{aligned}
\end{equation}

\noindent which has a maximum at $\rho_p=\frac{3}{4R_p}\sqrt{\frac{3\sigma_s}{\pi G}}$ and is independent of $Q_0$.

Due to the fact that the pericentre distance $q$ increases while $\omega_{\mathit{eq}}$ decreases with tidal evolution (Section \ref{model properties}), the condition $q_0\geq r_{\mathit{Roche,0}}$ is sufficient to avoid tidal disruption assuming that the properties of the object remain the same. 

\subsection{Summary}\label{method summary}

\begin{figure*}
\includegraphics[width=0.8\textwidth]{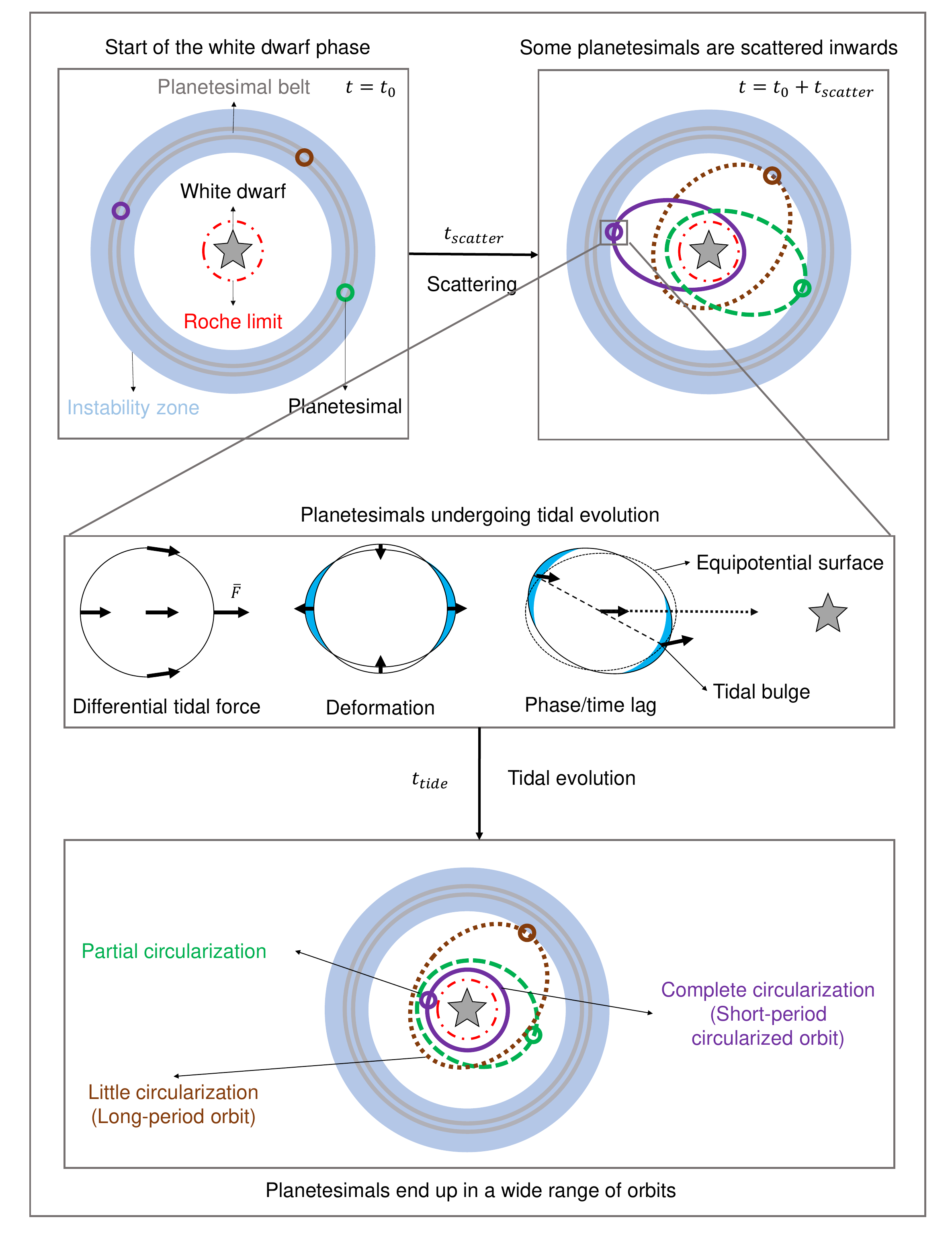}
\caption{A schematic diagram of planetesimals scattered towards a white dwarf that evolve onto shorter-period orbits under tides (face-on view, not to scale). Three snapshots in time are presented: 1. at the start of the white dwarf phase (upper-left panel), 2. when the planetesimal is initially scattered (upper-right panel), and 3. after a given tidal evolution time (lower panel) where planetesimals evolve onto a wide range of orbits. The stages of tidal evolution can be categorized as: complete circularization (purple, small $q_0$, short-period circular orbit), partial circularization (green, moderate $q_0$) and little circularization (brown, large $q_0$, long-period highly eccentric orbit). The middle panel is a sketch of the tidal effect: 1. the white dwarf exerts a differential tidal force on the planetesimal (left), 2. the planetesimal is deformed, leading to the formation of tidal bulges (middle) and 3. the lag of tidal bulges relative to the equipotential surface (right) leads to dissipation of orbital energy.}
\label{scattering plot}
\end{figure*}

\begin{table}
\begin{center}
\begin{tabular}{|c c|c| } 
 \hline
Planetesimal&$\rho_p\,(\mathrm{kg/m^3})$&3000 \\ 

&$R_p\,(\mathrm{km})$&100 \\ 

&$K_p\,(\mathrm{s})$&1000 \\ 

&$C_p$&$\frac{2}{5}$\\ 

&$\sigma_s\,(\mathrm{Pa})$&$10^{5}$ \\

\hline
White dwarf&$M_{*}\,(M_{\odot})$&0.6 \\ 
\hline
Orbit&$Q_0\,\rm (AU)$& 3 \\
& $\omega_{\mathit{p,0}}\,(\omega_{\mathit{eq,0}}(\epsilon_p=0))$ &1\\
& $\epsilon_{\mathit{p,0}}$ &0\\
& $P(q_0)$ & $P(q_0)\propto q_0^{\frac{1}{2}}$ \\
\hline
\end{tabular}
\end{center}
\caption{The fiducial set of parameters used in this work to describe a planetesimal-white dwarf system undergoing tidal evolution. The choice of $K_p$ is motivated by the tidal dissipation of typical rocky bodies \citep{1977RSPTA.287..545L,1997A&A...318..975N,2011A&A...535A..94B,2015A&A...584A..60C,2017CeMDA.129..509B,2020JGRE..12506312R,2020A&A...637A..78S, 2022AdGeo..63..231B}. The choices of $\rho_p$ and $\sigma_s$ are motivated by the physical properties of chondrites \citep{2019P&SS..165..148O,2020M&PS...55..962P}. The choice of white dwarf mass roughly corresponds to the peak of the observed white dwarf mass distribution \citep{2007MNRAS.375.1315K,2016MNRAS.461.2100T,2017ASPC..509..421K}.The initial apocentre of the planetesimals, $Q_0$, is set such their orbits did not enter the stellar envelope during the giant branches, i.e. slightly above the maximum size of the stellar envelopes of 1--3\,$M_{\odot}$ stars with solar metallicity based on MESA isochrones and stellar tracks (MIST, \citealp{2016ApJS..222....8D,2016ApJ...823..102C}). The initial spin and obliquity are arbitrarily chosen to be their equilibrium values as they have little effect on the semi-major axis and eccentricity evolution (see Section \ref{tidal evolution track}).}
\label{tidal parameters table}
\end{table}

\begin{figure}
\begin{center}
\includegraphics[width=0.3\textwidth]{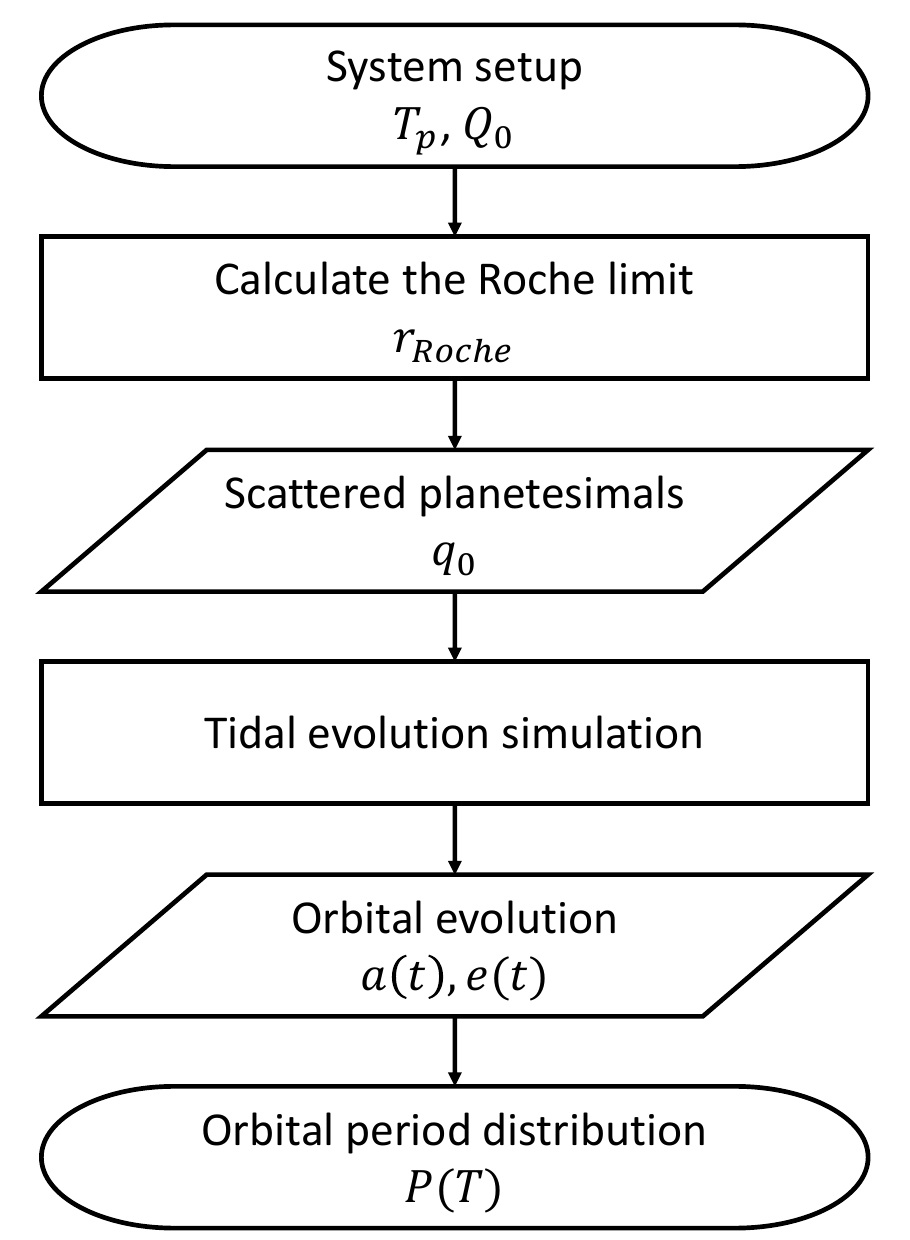}
\caption{A flow chart presenting the method described in Section \ref{method}: 1. initial setup of a white dwarf-planetesimal system (two-body system properties $T_p$ and initial apocentre distance $Q_0$), 2. compute the Roche limit ($r_{\mathit{Roche}}$), 3. imitate the scattering process via an initial pericentre ($q_0$) distribution whose lower limit is set by the Roche limit,  4. simulate the tidal evolution, 5. obtain the time evolution of orbital parameters ($a(t)$, $e(t)$), and 6. compute the resultant orbital period distribution.}
\label{method plot}
\end{center}
\end{figure}   

The physical processes considered in this study are summarized in Fig.\ref{scattering plot}: planetesimals scattered to pericentre distances narrowly avoiding tidal disruption experience strong tidal force, inducing tidal bulges that lag behind the equipotential surface, damping orbital energy, bringing these planetesimals to a wide range of orbits: 1. short-period circularized orbits (purple), 2. partially circularized orbits (green), 3. long-period orbits nearly unaffected by tidal effect (brown). 

For identical two-body system properties, tidal evolution rate increases rapidly with the decreasing initial pericentre distance (Section \ref{model properties}). The initial pericentre distance $q_0$ and initial apocentre distance $Q_0$ of the planetesimal after leaving the instability zone are shaped by the white dwarf planetary system, which comes with a wide range of architectures. Instead of simulating the dynamical processes delivering planetesimals close to the white dwarf, we introduce a power law probability density function (PDF) of $q_0$ of the planetesimal to imitate the effect of scattering: $P(q_0\geq r_{\mathit{Roche,0}})\propto q_0^{\alpha}$. We obtain a population of $q_0$ that the planetesimal is scattered to according to the PDF using the Metropolis algorithm with a sample size of $10^{5}$. With the lower limit of $q_0$ set by the Roche limit, we set the upper limit to 0.012\,AU, above which the planetesimal's orbital period can hardly decay below 1\,yr within 1\,Gyr, adding difficulties to observations. The default value of $\alpha$ chosen for this study is $\alpha=\frac{1}{2}$. We will discuss the choice of this PDF and the effect of different $\alpha$ in Section \ref{free parameters}, and show that the resultant orbital period distribution is qualitatively insensitive to the assumed initial pericentre distribution. $Q_0$ physically represents the inner edge of the scattering zone, which may vary a lot among different systems. For planets as perturbers, $Q_0$ depends on planet-planet interactions (e.g., scattering, resonances) and planet-star interactions (e.g., tidal interaction, common-envelope evolution), and is not necessarily equivalent to the semi-major axis of a planet would have at the end of the asymptotic giant branch assuming adiabatic orbital expansion under stellar mass loss. We will discuss the effect of $Q_0$ on the circularization timescale and the orbital period distribution in Section \ref{circularization timescale} and Section \ref{free parameters}.

The default setup of the white dwarf-planetesimal system is listed in Table \ref{tidal parameters table}, unless otherwise stated (see the caption for the motivation of choices). The properties of exoplanetary bodies, especially the tidal dissipation efficiency quantified by $K_p=3k_2\Delta t$, are poorly constrained, and may be altered during tidal evolution. Knowing the tidal evolution time and the tidal evolution stage relative to the initial condition ($a(t)$, $e(t)$, $q_0$, $Q_0$) is insufficient to disentangle the parameters within $T_p$ (Eq.\ref{Tp equation}). Furthermore, the constituents of $T_p$ are not necessarily independent variables. Hence, it is more appropriate to consider $T_p$ as a whole. However, we will still investigate the effect of varying the individual constituents of $T_p$ independently (e.g., density, radius) for illustrative purpose. On the other hand, as is mentioned in Section \ref{model properties}, planetesimals starting with same $q_0$ and $Q_0$ would have identical orbital parameters at identical $T_pt$. Therefore, for a realistic $T_p$ that is $k$ times the assumed value in this study, one can obtain the realistic counterparts of our simulation results by substituting $\frac{t}{k}$ to $t$.

With the initial orbital parameters and the properties of the planetesimal-white dwarf system, one can simulate the orbital evolution of a planetesimal. 

The method is summarized in Fig.\ref{method plot}:

\begin{enumerate}
    \item We setup the properties of the two-body system.
    \item We compute the Roche limit as the lower limit of initial pericentre distance of planetesimals.
    \item We use the Metropolis algorithm to generate a sample of initial pericentre distance outside the Roche limit according to the power law probability density function, $P(q_0\geq r_{\mathit{Roche,0}})\propto q_0^{\alpha}$.
    \item We simulate the tidal evolution of sample planetesimal-white dwarf systems by linearly interpolating within a pre-simulated grid of tidal evolution tracks for a range of initial pericentre distances.
    \item We obtain the orbital evolution of planetesimals around white dwarfs, from which we deduce the orbital period distribution.
\end{enumerate}

\section{Results}\label{results}

\subsection{Tidal evolution under the CTL model}\label{tidal evolution under the CTL model}

To obtain the orbital period of a planetesimal at a given snapshot in time, we need to model its orbital evolution. In this section we model the tidal evolution tracks of sample systems, comparing them to the analytical analysis in \ref{model properties} (\ref{tidal evolution track}), and investigate the parameter-dependence of tidal circularization timescale (\ref{circularization timescale}). 

\subsubsection{Tidal evolution tracks}\label{tidal evolution track}

\begin{figure}
\includegraphics[width=0.45\textwidth]{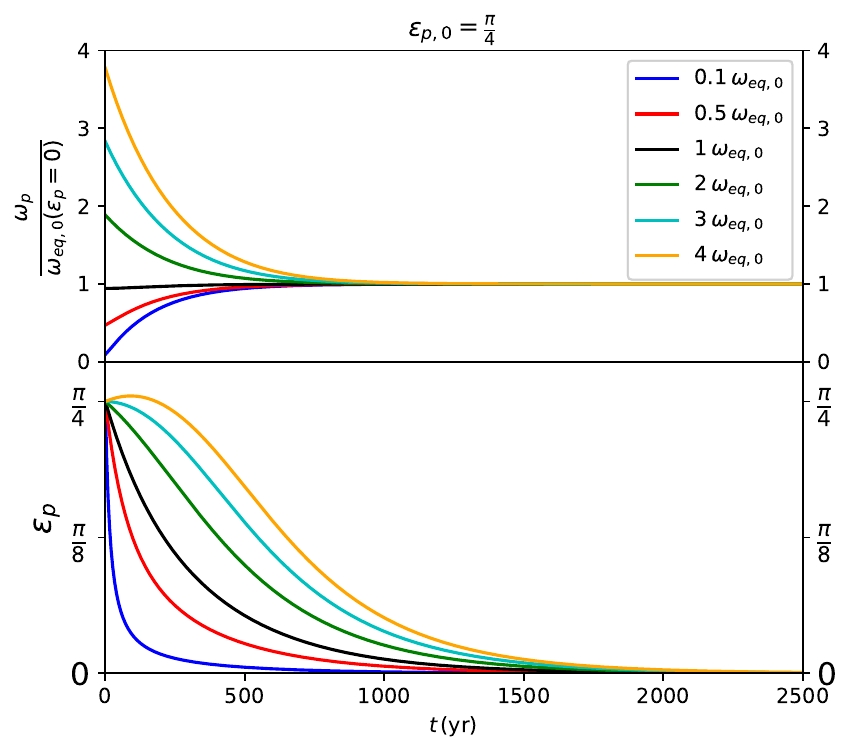}
\caption{The tidal evolution of the spin ($\omega_p$, upper panel) and obliquity ($\epsilon_p$, lower panel) of the planetesimal under the CTL model for an initial obliquity of $\frac{\pi}{4}$. We choose $q_0=0.01\,\rm AU$. Each colour corresponds to a different initial spin expressed as a multiple of the initial equilibrium spin with 0 obliquity $\omega_{\mathit{eq,0}}(\epsilon_{\mathit{p}}=0)$. $a$ and $e$ remain identical on the timescale of this plot. Other choice of free parameters are listed in Table \ref{tidal parameters table}.}
\label{tidal omega i plot}
\end{figure}

\begin{figure}
\includegraphics[width=0.45\textwidth]{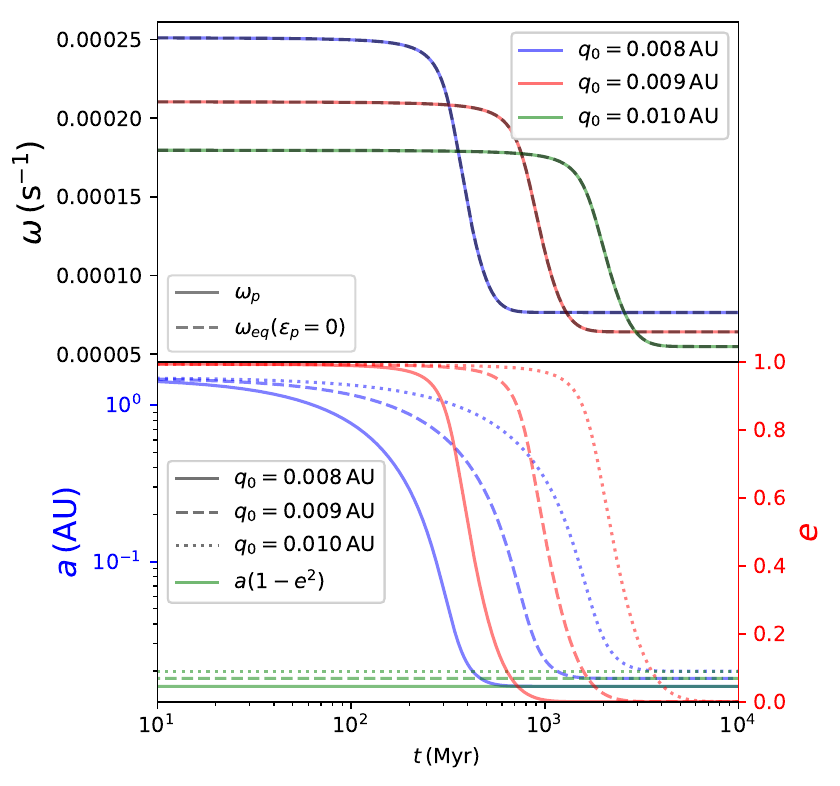}
\caption{The tidal evolution of $\omega_p$ (solid lines), $\omega_{\mathit{eq}}$ (black dashed lines) in the upper panel, which perfectly overlap with each other, and the evolution of $a$ (blue lines, left axis), and $e$ (red lines, right axis) in the lower panel, for $q_0=0.008, 0.009, 0.010\,\rm AU$. The green lines are $a(1-e^2)$ (left axis) for the corresponding $q_0$. Other free parameters are listed in Table \ref{tidal parameters table}.}
\label{tidal a e weq plot}
\end{figure}

In Fig.\ref{tidal omega i plot}, we plot the evolution of spin, $\omega_p$ (upper panel) and obliquity, $\epsilon_p$ (lower panel) at $q_0=0.01\,\rm AU$ for different initial spins (expressed as the multiple of $\omega_{\mathit{eq,0}}(\epsilon_p=0)$) over 2500\,yr. The key features are summarized below:

\begin{itemize}
    \item Spin and obliquity converge to their equilibrium values on similar timescales ($\sim 1000 \,\rm yr$), within which semi-major axis and eccentricity, which decay on much longer timescales ($\sim 1\,\rm Gyr$, see Fig.\ref{tidal a e weq plot}, lower panel) remain unchanged.
    \item Spin and obliquity converge to the common equilibrium values, $\omega_{\mathit{eq,0}}(\epsilon_{p}=0)$ and 0, respectively, insensitive to the initial spin and obliquity.
    \item A smaller initial spin rate corresponds to a faster obliquity decay (lower panel, from the blue line to the orange line).
    \item For large initial spin, there exists temporary increasing obliquity opposite to the general declining trend (lower panel, orange line).

\end{itemize}

These features are consistent with the deduced properties of the CTL model. The key deduction from spin and obliquity evolution is that the choice of initial spin and obliquity has little effect on the tidal evolution in $a$--$e$ space.

In Fig.\ref{tidal a e weq plot}, we plot the evolution of $\omega_p$ (solid lines) and $\omega_{\mathit{eq}}$ (black dashed lines) for 3 different $q_0$: 0.008\,AU (blue), 0.009\,AU (red) and 0.010\,AU (green) for 10\,Gyr in the upper panel, together with the corresponding evolution of $a$ and $e$ in the lower panel. In the upper panel, the persistent overlap of the solid lines ($\omega_p$) and the dashed lines ($\omega_{\mathit{eq}}(\epsilon_p=0)$) indicates that a planetesimal effectively remains in the pseudo-synchronization and spin-orbit alignment state when considering the evolution in $a$ and $e$. 

The green lines in the lower panel representing $a(1-e^2)$ (left axis) are horizontal, and perfectly overlap with the value that the blue lines ($a$) converge to, indicating conservation of orbital angular momentum throughout tidal evolution, consistent with Section \ref{model properties}. In other words, the initial orbital parameters, $q_0$ and $Q_0$, determine the $a$--$e$ relation during tidal circularization and the semi-major axis after circularization.

On the other hand, as is shown in the lower panel of Fig.\ref{tidal a e weq plot}, the convergence of eccentricity always lags behind that of the semi-major axis, which is consistent with conservation of orbital angular momentum (Eq.\ref{a-e relation from constant L}): $\frac{da}{a}=\frac{2e^2}{1-e^2}\frac{de}{e}$, such that $\frac{da}{a}\gg\frac{de}{e}$ at large $e$ and that $\frac{da}{a}\ll\frac{de}{e}$ at small $e$ (or, note that $|\frac{da}{dt}|$ decays monotonically towards $e\rightarrow 0$ to 0, while $|\frac{de}{dt}|$ decays from its maximum, at $e\approx 0.658$, towards both $e\rightarrow 0$ and $e\rightarrow 1$, to 0, Section \ref{model properties}). Consequently, an orbit is effectively circularized in terms of its orbital period when the eccentricity is small ($e\lesssim 0.01$ in Fig.\ref{tidal a e weq plot}), and it takes much longer to reach the true circularization point $e=0$.

With the increase in $q_0$ from 0.008\,AU to 0.010\,AU, the time required for the semi-major axis (blue lines) to converge increases rapidly from $\sim 700 \,\rm Myr$ to $\sim 4000\,\rm Myr$, roughly consistent with $\tau_{\mathit{cir}}\sim \left|\frac{a}{\dot{a}}\right|_0\propto q_0^{7.5}$ (Appendix B). Furthermore, at same $t$, the orbital period difference of two planetesimals starting at different $q_0$ initially increases and then decays during tidal evolution, reaching a constant after both planetesimals are circularized.

\subsubsection{Tidal cirularization timescale}
\label{circularization timescale}

\begin{figure*}
\includegraphics[width=0.8\textwidth]{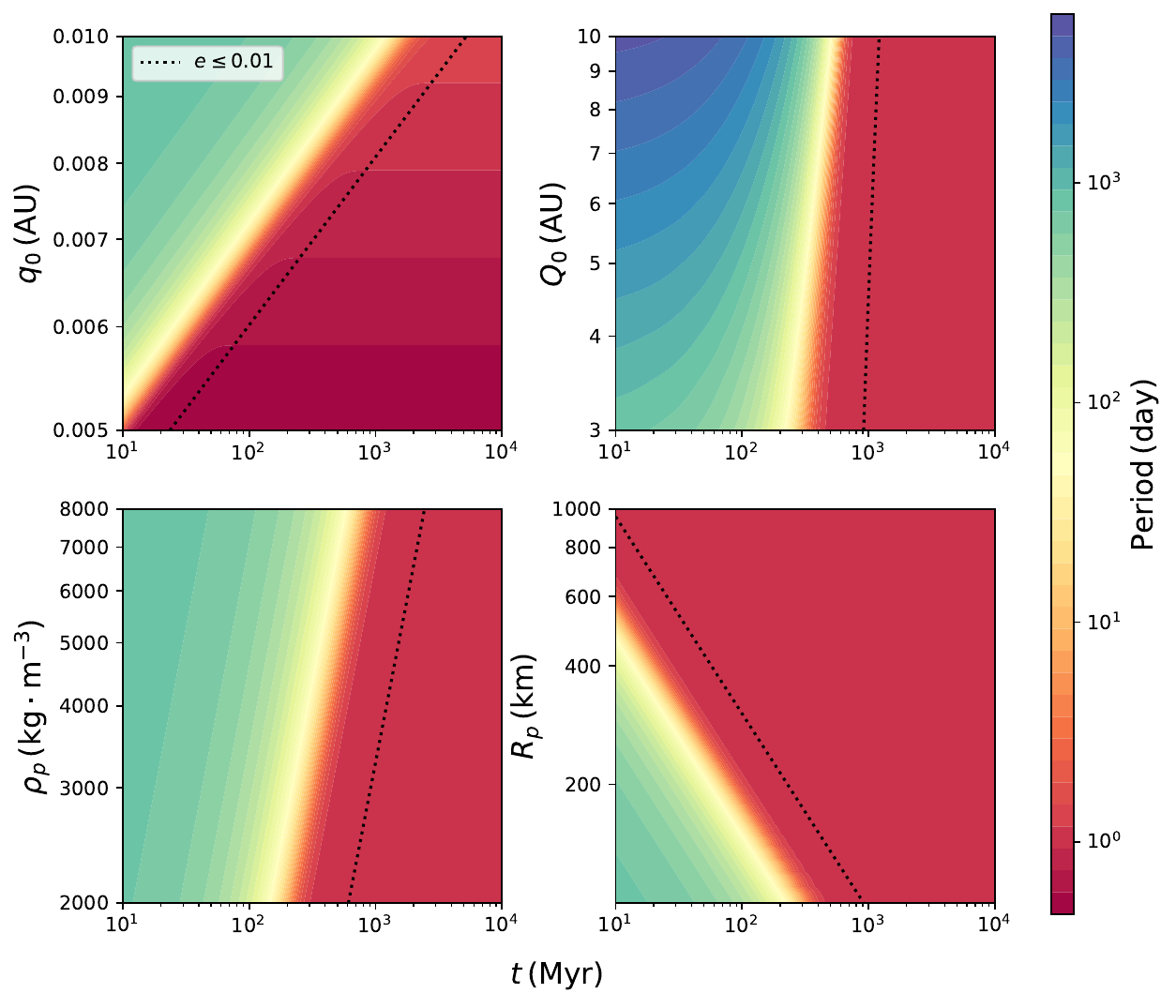}
\caption{The contour plots of the orbital period of the planetesimal in $t$--$q_0$ space (upper-left panel), as well as its counterparts in $t$--$Q_0$ space (upper-right panel), $t$--$\rho_p$ space (lower-left panel) and $t$--$R_p$ space (lower-right panel) at a fixed $q_0=0.008 \,\rm AU$. The axes are all on the logarithmic scale. Other free parameters are identical to those in Table \ref{tidal parameters table}. The dotted lines are the circularization lines, at the right of which $e\leq 0.01$ is satisfied.}
\label{period time contour}
\end{figure*}

\begin{table}
\begin{center}
\begin{tabular}{|c|c|c|c| } 
 \hline
$\frac{\partial\log\tau_{\mathit{cir}}}{\partial\log q_0}$ &$\frac{\partial\log\tau_{\mathit{cir}}}{\partial\log Q_0}$ & $\frac{\partial\log\tau_{\mathit{cir}}}{\partial\log \rho_p}$ &$\frac{\partial\log\tau_{\mathit{cir}}}{\partial\log R_p}$\\

\hline

7.8 & 0.2 & 1.0 & -2.0\\ 

\hline 

\end{tabular}
\end{center}
\caption{The gradient of circularization lines on the logarithmic scale in Fig.\ref{period time contour}.}
\label{circularization table}
\end{table}

If planetesimals are to evolve onto (nearly) circular orbits around white dwarfs due to tides, this evolution must occur well within the white dwarf cooling age. In this section we simulate the variation in circularization timescale ($\tau_{\mathit{cir}}$) in response to the physical properties of the planetesimal and the initial orbital parameters. We consider planetesimal size and density as independent variables for demonstration only, which is not necessarily realistic.

In Fig.\ref{period time contour}, we plot the orbital period evolution of the planetesimal in $t$--$q_0$ (upper-left panel), $t$--$Q_0$ (upper-right panel), $t$--$\rho_p$ (lower-left panel) and $t$--$R_p$ (lower-right panel) space. The dotted lines are the circularization lines under the condition $e\leq 0.01$, at the right of which the planetesimal's orbit is considered to be effectively circularized. We do not use $e=0$ as the condition of circularization because a nearly circularized orbit is non-distinguishable from, but is reached much earlier than its circularized counterpart, as mentioned in Section \ref{tidal evolution track}. The gradients of circularization lines on the logarithmic scale are summarized in Table \ref{circularization table}. We find out that these gradients are not sensitive to the numerical condition of circularization (for instance, the computed gradients are almost identical when setting the threshold to be $e\leq 10^{-4}$ instead) and other free parameters (for instance, $Q_0$ in the upper-left panel, $q_0$ in upper-right and lower panels).

Clearly, the gradient of the circularization line in the upper-right panel, corresponding to $\frac{\partial\log\tau_{\mathit{cir}}}{\partial\log q_0}\approx 7.8$, far exceeding the magnitude of others, indicating that the initial pericentre distance $q_0$ dominates the circularization timescale over other free parameters. The empirical relation $\tau_{\mathit{cir}}\propto q_0^{7.8}$ (with the scaling insensitive to a wide range of $Q_0$ and $T_p$) is roughly consistent with the analytically computed circularization timescale $\left|\frac{a}{\dot{a}}\right|_0\propto q_0^{7.5}$ (the difference originates from the fact that the analytical expression is obtained by expanding $\frac{a}{\dot{a}}$ at $t=0$, which becomes less accurate as the orbital parameters deviate from their initial values under tidal evolution). At any snapshot in time, the orbital period of the planetesimal decreases monotonically with $q_0$. If a planetesimal starts at a smaller pericentre distance, it would circularize much more rapidly and to a shorter orbital period.

The Roche limit of the planetesimal with the default properties (Table \ref{tidal parameters table}) is $0.00456\,\rm AU$, corresponding to a minimum circularization timescale of $\sim 10 \,\rm Myr$. Therefore, for a white dwarf with a cooling age of 1\,Gyr, there may exist tidally circularized planetesimals unless the tidal dissipation of exoplanetary bodies quantified by $T_p$ (Eq.\ref{Tp equation}) is $\gtrsim 100$ times smaller than our fiducial value (Table \ref{tidal parameters table}). In a regime where $\rho_p$ can be considered as an independent variable, and when the self-gravity far exceeds the material strength, Eq.\ref{roche limit approximated equation} is reduced to $r_{\mathit{Roche}}\propto \rho_p^{-\frac{1}{3}}$. Therefore, the minimum circularization timescale ($\tau_{\mathit{cir}}\propto r_{\mathit{Roche}}^{7.8}\rho_p$) is roughly proportional to $\rho_p^{-1.6}$, such that increasing planetesimal density from $3000\,\mathrm{kg/m^3}$ to $4000\,\mathrm{kg/m^3}$ reduce the minimum circularization timescale to $\sim 6 \rm\,Myr$. 

Among the free parameters investigated in this section, the dependence of $\tau_{\mathit{cir}}$ on $Q_0$ is the weakest, as is shown in the nearly vertical circularization lines in the upper-right panel of Fig.\ref{period time contour} corresponding to $\tau_{\mathit{cir}}\propto Q_0^{0.2}$. However, the orbital period of partially circularized planetesimals (at the left of the circularization lines) are increasingly affected by $Q_0$ at an earlier tidal evolution stage, where a larger $Q_0$ corresponds to a $a$ and hence larger $T(t)$.

The delayed circularization with the increase in $\rho_p$ in the lower-left panel, with $\tau_{\mathit{cir}}\propto \rho_p$, and the accelerated tidal circularization with increasing planetesimal size in the lower-right panel, with $\tau_{\mathit{cir}}\propto R_p^{-2}$, are consistent with the fact that $\dot{a}$ and $\dot{e}$ are proportional to $ T_p\propto R_p^2\rho_p^{-1}$.

Combining the scaling relations above, we obtain an empirical scaling relation for the circularization timescale of planetesimals around white dwarfs under the CTL model:

\begin{equation}\label{t_cir scaling}
\begin{aligned}
\tau_{\mathit{cir}}&\approx 5000\frac{1000\,\mathrm{s}}{K_p}\left(\frac{0.6\, M_{\odot}}{M_*}\frac{100\,\mathrm{km}}{R_p}\right)^2\frac{\rho_p}{3000\,\mathrm{kg/m^3}}\\& \times\left(\frac{q_0}{0.01\,\rm AU}\right)^{7.8}\left(\frac{Q_0}{3\,\rm AU}\right)^{0.2}\,\mathrm{Myr}.     
\end{aligned}
\end{equation}

\subsection{Probability distribution of orbital period}\label{period distribution}

\begin{figure*}
\includegraphics[width=0.8\textwidth]{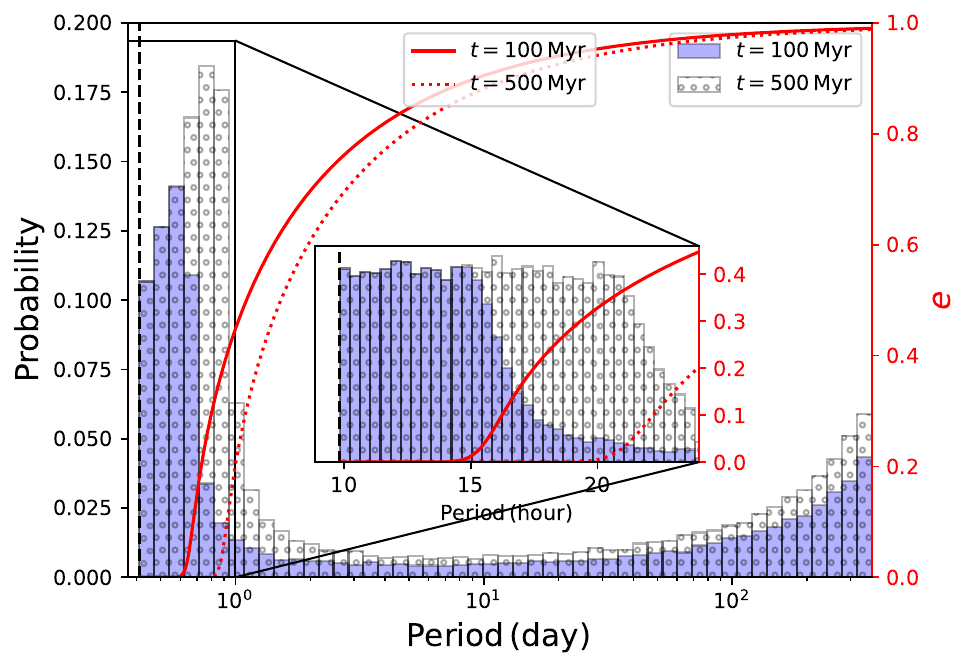}
\caption{The simulated normalised probability distribution of the planetesimal's orbital period (left axes, histogram) and eccentricity (right axes, lines) within 1\,yr on the logarithmic scale, at a tidal evolution time of 100\,Myr (blue histogram, solid line) and 500\,Myr (hatched histogram, dotted line), together with the zoom-in plot for orbital period less than 1\,day on the linear scale. The free parameters are listed in Table \ref{tidal parameters table}.}
\label{sample_t_subplot_log}
\end{figure*}

\begin{figure}
\begin{center}
\includegraphics[width=0.45\textwidth]{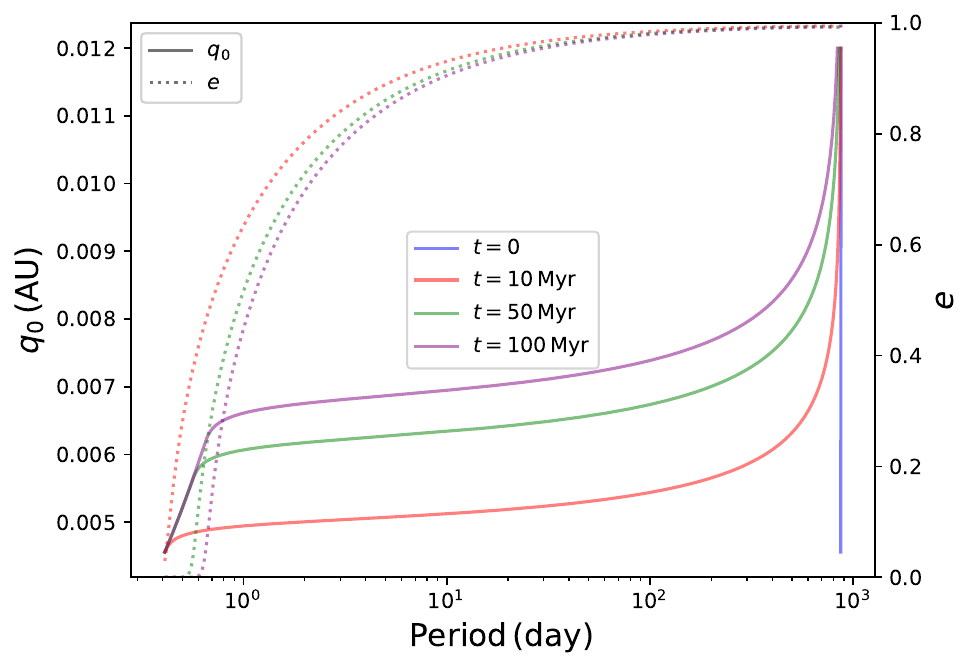}
\caption{The orbital period and orbital eccentricity a planetesimal scattered to $q_0$ would have at 4 snapshots in time for a range of $q_0$. The free parameters are identical to those in Table \ref{tidal parameters table}.}
\label{q0 period plot}    
\end{center}
\end{figure}

This section predicts the probability distribution of the orbital period for a planetesimal evolving under tidal interactions (which can be equivalently interpreted as the orbital period distribution of a population of planetesimals with identical properties and tidal evolution times). The orbital (transiting) period is the key observable to compare with. We note here that the probability distribution of a given planetesimal (given planetesimal population) is different if plotted in terms of the planetesimal's orbital period rather than semi-major axis, on linear rather than logarithmic scale (see Appendix C). 

We investigate the probability distribution in the period space and on the logarithmic scale for orbital periods $\leq$ 1\,year, since it is more natural to compare between $P(0.9\,\rm day\leq T\leq 1.1\,\rm day)$ to $P(90\,\rm day\leq T\leq 110\,\rm day)$, instead of comparing between $P(0.9\,\rm day\leq T\leq 1.1\,\rm day)$ to $P(99.9\,\rm day \leq T \leq 100.1\, \rm day)$.

In Fig.\ref{sample_t_subplot_log}, we show 2 snapshots in tidal evolution time: 100\,Myr and 500\,Myr of the probability distribution of planetesimal's orbital period limited to $T\leq 1\,\rm yr$ (histogram, left axis) on the logarithmic scale, with the zoom-in plot for the short period orbits ($T\leq 1\,\rm day$) in the linear space. Note that $T_p$ and $t$ 
are entangled such that the distribution at a tidal evolution time of 100\,Myr can be equivalently understood as a distribution at $t=500\,\rm Myr$ but for a planetesimal with $T_p$ that is 5 times weaker (tidal evolution is simultaneous in $T_pt$ space, \ref{model properties}). The probability distribution is normalised such that at $t=100\,\rm Myr$, the probability within the cut-off period, 1\,yr, adds up to 1. The dashed line is the minimum circularization period ($P_{\mathit{cir}}(r_{\mathit{Roche}})$) computed from Eq.\ref{a-e relation from constant L} and Eq.\ref{solve roche limit equation}. The solid and dotted lines are the orbital eccentricity a planetesimal would have at current orbital period, at a tidal evolution time of 100\,Myr and 500\,Myr, respectively.

The key features of Fig.\ref{sample_t_subplot_log} are summarized below:

\begin{enumerate}
    \item A peak of the probability distribution presents near the transition between (nearly) circularized and partially circularized orbits ($\sim 10\,\rm hr$--$1\,\rm day$): an enhanced probability for the planetesimal to reside on a (nearly) circularized orbit. 
    \item There is a probability valley after the peak: low probability for the planetesimal to reside in a partially circularized orbit with an orbital period between $\sim 1\,\rm day$ and $\sim 100\,\rm day$.
    \item There is an increasing tail towards longer-period highly eccentric orbits ($\gtrsim 100 \,\rm day$).
    \item As the time of tidal evolution increases (the hatched histogram versus the blue histogram), the peak of the probability distribution shifts towards longer period and the contributions from the (nearly) circularized orbits to the probability distribution increases.
    \item The probability distribution of the (nearly) circularized orbits is uniform in the linear space (zoom-in plot) and hence linear in the logarithmic space.
\end{enumerate}

Notably, the pile-up at short-period nearly circularized orbits and long-period highly eccentric orbits ((i), (ii) and (iii)) is a natural consequence of the tidal model and the utilization of logarithmic scale, rather than the initial pericentre distribution. This is shown in Fig.\ref{q0 period plot}, which illustrates that a given bin-width in logarithmic period space (fixed $\frac{\Delta T}{T}$) covers a wider range of $q_0$ towards short-period nearly circularized orbits and long-period highly eccentric orbits under tidal evolution. Fig.\ref{q0 period plot} is equivalent to a cumulative distribution for a uniform distribution in $q_0$ and the orbital period distributions of tidally evolved planetesimals are strongly shaped by the pile-up illustrated in Fig.\ref{q0 period plot}. The physical origin of this pile-up is briefly explained in Appendix C, from which we can deduce that, as long as the realistic tidal evolution remains qualitatively similar to the predictions of the CTL model in terms of:

\begin{itemize}
    \item semi-major axis decay generally decelerates along the tidal evolution track, converging to 0 at $e=0$,
    \item a planetesimal closer to the white dwarf at its pericentre undergoes more rapid tidal evolution,
\end{itemize}

\noindent the qualitative features (i) (ii) and (iii) of the orbital period distribution should persist, regardless of the assumptions regarding the initial pericentres (see Section \ref{free parameters}).

On the other hand, the tidal model and the planetesimal population (e.g., initial pericentre distribution, physical properties) can affect the orbital period distribution quantitatively. The inner edge of the orbital period distribution is an indication of the Roche limit, constraining the physical properties such as density and tensile strength of the planetesimal. The position of the peak at short-period, indicating the transition from near-circular to eccentric orbits not only indicates the tidal evolution stage of the system but also illustrates the maximum pericentre distance where a planetesimal scattered to can be tidally circularized within a given timescale. The peak for nearly circularized planetesimals, if presenting at a larger orbital period, indicating a later tidal evolution stage under the CTL model, i.e., longer tidal evolution time/stronger tidal interactions ((iv)), because CTL model predicts that a planetesimal starting with a larger $q_0$ circularizes slower to a longer orbital period. The time evolution of the orbital period distribution is affected by the dependence of tidal model on the initial pericentre distance. The orbital period distribution of nearly circularized orbit, as well as the fraction of nearly circularized to partially circularized orbits, is related to the initial pericentre distribution. For instance, our choice of PDF $P(q_0)\propto q_0^{\frac{1}{2}}$ corresponds to an uniform orbital period distribution in the linear space (linear in logarithmic period space) for nearly circularized planetesimals ((v), Appendix C). To summarize, the qualitative features of the orbital period distribution: peak at short-period nearly circularized orbit and an increasing tail towards long-period highly eccentric orbits persist for any tidal evolution that is qualitatively similar to the CTL model, while the quantitative details of the distribution probes the tidal model, as well as the population of planetesimals undergoing tidal evolution.

\section{Discussion}\label{discussion}

\begin{figure*}
\includegraphics[width=0.8\textwidth]{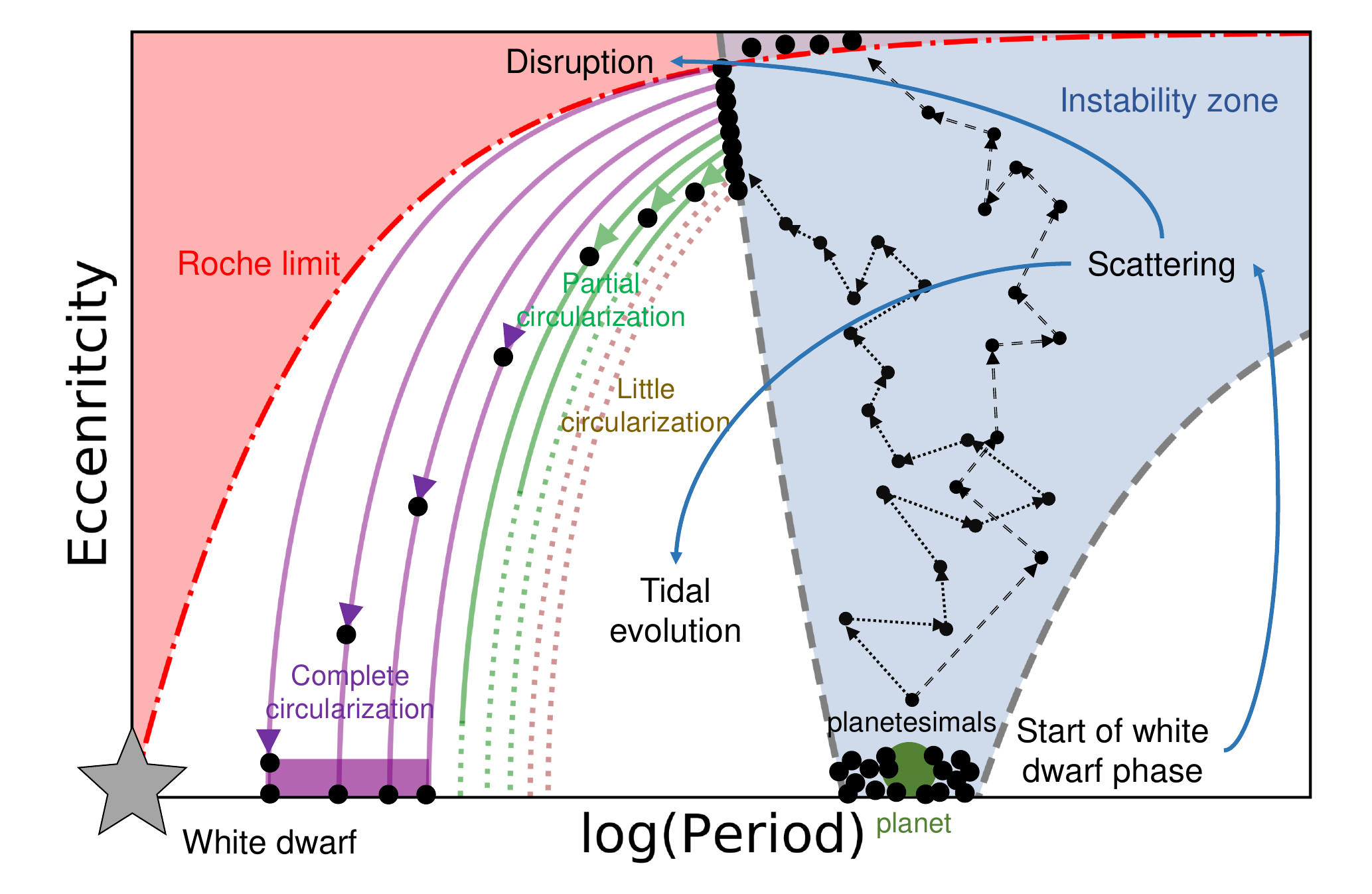}
\caption{A schematic illustrating two potential pathways to white dwarf pollution from planetesimals scattered inwards by planets. Planetesimals that are scattered interior to the Roche limit are tidally disrupted and accreted. Planetesimals scattered outside the Roche limit evolve under tides, potentially leading to a wide range of orbits around the white dwarf.}
\label{scattering plot 1}
\end{figure*}

This paper studies an alternate pathway for planetesimals scattered close to the white dwarf. While planetesimals scattered interior to the Roche limit are tidally disrupted and accreted by the white dwarf, planetesimals scattered just outside the Roche limit may undergo tidal circularization. These tidally evolved planetesimals potentially span a wide range of orbital periods at a given snapshot in time (Fig.\ref{scattering plot 1}). These planetesimals may be responsible for transits observed around white dwarfs. 

The observed light curves for the white dwarfs with photometric variability are not consistent with the transit of an intact body, but do contain periodic signatures that could be explained by active dust production associated with planetary bodies \citep{2015Natur.526..546V,2017MNRAS.465.1008V,2020ApJ...893..166D}. If the observed periods correspond to the orbits of planetesimals, this work presents a mechanism for planetesimals to arrive on short period orbits. Mechanisms that allow planetesimals on short period orbits to produce photometrical variability in white dwarfs will be the subject of Paper II in this series (see \citealp{2015Natur.526..546V,2017MNRAS.465.1008V,2020ApJ...893..166D} for existing models). 

Planetesimals that are tidally evolved outside the Roche limit and their counterparts tidally disrupted inside the Roche limit may coexist, with the ratio between them potentially varying in different systems and different evolutionary stages. The key test of this model is the distribution of orbital periods in the population of white dwarfs with observed transits. In this section, we firstly discuss the limitations of the model that affect the predictions made regarding the orbital period distributions of planetesimals shaped by tidal evolution, starting with the tidal model (Section \ref{tidal model discussion}) and the choice of the initial conditions, which dominate the subsequent tidal evolution (Section \ref{free parameters}). Our model predictions are then compared to the current observations (Section \ref{period distribution discussion} and \ref{roche limit discussion}), before the implications of an alternative pathway for planetesimals to evolve in white dwarf planetary systems are discussed (Section \ref{implication}).

\subsection{Tidal model}\label{tidal model discussion}

Although the details of the prediction for the distribution of orbital periods of planetesimals orbiting white dwarfs depends on how white dwarf-planetesimal systems evolve under tide, the general properties of the distribution, most notably the pile-up of planetesimals on short-period (nearly) circularized orbits (which originates from the decelerating decay rate of orbital period during tidal circularization) is robust. Here we discuss the limitations in our understanding of how tides act on planetesimals and how these influence the results presented in this work. 

In this study, we utilize the CTL model, which assumes an equilibrium tidal bulge lags behind the equipotential surface for a constant time. The CTL model is a simplified tidal model equivalent to a rheology where the induced tidal response is proportional to the corresponding tidal forcing frequency, such that the summed tidal dissipation efficiency of all forcing frequencies can be characterized by a single proportionality constant (e.g., $K_p$ in this study). On one hand, high-eccentricity tidal evolution may be fundamentally different from its low-eccentricity counterparts because high-eccentricity tidal interactions mainly occur near the pericentre where an orbital averaged description as is the case for the CTL model may become inaccurate, and tidal dissipation may originate from different physics as is described by the CTL model (for instance, excitation and dissipation of oscillation modes for gas giants and stars) \citep{1977ApJ...213..183P,2018ApJ...854...44M,2019MNRAS.484.5645V,2022ApJ...931...11G,2022ApJ...931...10R, 2019MNRAS.484.5645V, 2020MNRAS.492.6059V}.

On the other hand, for realistic rheologies, the dependence of tidal response on the forcing frequency may no longer lie in the linear regime as is the case for the CTL model \citep{2013ApJ...764...27M,2014Icar..241...26N,2014MNRAS.438.1526S,2019CeMDA.131...30B,2019MNRAS.486.3831V,2023ApJ...943L..13V}. Furthermore, a stable pseudo-synchronization state may not present at all orbital parameters \citep{2013ApJ...764...27M,2014MNRAS.438.1526S,2020MNRAS.498.4005O,2020MNRAS.496.3767V}, making the Roche limit more uncertain and tidally-induced rotational fission possible. Furthermore, thermal evolution of the planetesimal under tidal heating may alter its rheology, and hence the distribution of tidal energy inside the body, tidal response, as well as the Roche limit \citep{2005Icar..177..534T,2013Icar..223..308B,2015E&PSL.427...74Z,2024ApJ...961...22S}, adding complexities to the physical picture, for instance, runaway melting (tidal response increases with the degree of melting \citealp{2024ApJ...961...22S}), thermal regulation (tidal response decreases with the degree of melting \citealp{2015E&PSL.427...74Z}), re-triggered tidal disruption (weakening ultimate tensile strength, increasing degree of deformation, see \ref{roche limit discussion}) and thermal destruction \citep{2020MNRAS.492.6059V}. Notably, although tidal response (for instance, $T_p$ in the CTL model) and tidal evolution time are degenerate in terms of the orbital parameters (for instance, orbital period), the thermal history is potentially non-degenerate (rapid heating versus slow heating). A coupled thermal-tidal evolution model is required to properly model the tidal circularization of planetesimals around white dwarfs properly, which is beyond the scope of this study.

The most profound effect of the uncertainties in the tidal model is whether the tidal circularization timescale for planetesimals are so long that planetesimals can hardly circularize under tides within the cooling age of the white dwarfs. If tidal circularization is impossible throughout the allowed parameter space, the orbital period distribution in Fig.\ref{sample_t_subplot_log} is truncated before the orbital eccentricity approaches 0, such that the probability distribution may increase monotonically towards longer orbital period and the peak towards the short period (nearly) circularized orbits does not present (Appendix D). Conversely, at a given tidal evolution time, as long as there exists a parameter space where a planetesimal can become nearly tidally circularized, we do not expect a qualitative difference in the observed orbital period distribution from Fig.\ref{sample_t_subplot_log}, whose general features are shaped by the qualitative behaviours of the tidal model (see Section \ref{period distribution}), although the strength of the features, for instance, the height and position of the peak at nearly circularized orbits, can vary.

\subsection{Population of scattered planetesimals}\label{free parameters}

\begin{figure*}
\includegraphics[width=0.8\textwidth]{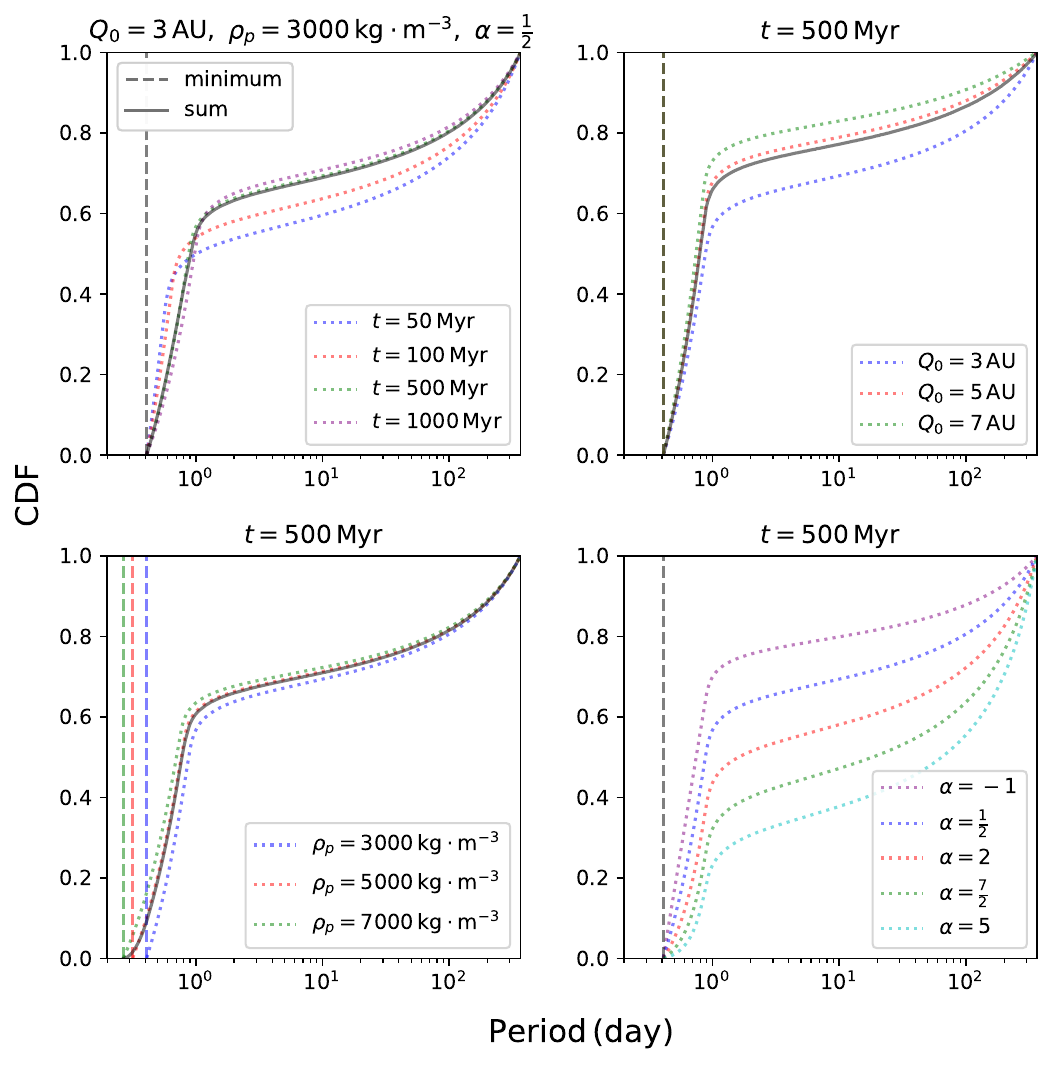}
\caption{The cumulative distribution of planetesimal's period within 1\,year for different evolution time (upper-left panel), initial apocentre distance (upper-right panel), density (lower-left panel) and initial pericentre distace distribution (lower-right panel). The default free parameters are identical to those in Table \ref{tidal parameters table} unless otherwise stated. The dashed lines are the minimum circularization period obtained from Eq.\ref{a-e relation from constant L}. The black solid lines are the CDFs obtained by summing the distribution for different tidal evolution time, initial apocentre distance and density uniformly within the range of the legend of each subplot.}
\label{CDF plot}
\end{figure*}

Given the large potential range in the ways that planetesimals can be perturbed inwards from an outer planetary system, it is impossible to constrain the exact population of scattered planetesimals. The population of scattered planetesimal is an important input for the model here, but we will show that the resultant orbital period distribution is qualitatively insensitive to the relevant free parameters. 

The assumed free parameters associated with the scattered planetesimal population around the white dwarf are: 

\begin{itemize}
    \item initial pericentre distance distribution (the likelihood and time taken of being scattered to a pericentre distance),
    \item initial apocentre distance (inner edge of the instability zone),
    \item property of white dwarf-planetesimal systems ($T_p$ and $r_{\mathit{Roche}}$),
    \item tidal evolution time (the difference between cooling timescale of the white dwarf and scattering timescale, degenerate with $T_p$).
\end{itemize}

In this work we arbitrarily choose a power law PDF to describe the initial pericentre distribution, motivated by numerical simulations of scattering \citep{2024MNRAS.527.11664}. We also note that a realistic initial pericentre distribution can be approximated as a power series via Taylor expansion at small $q_0$.  By investigating the resultant orbital period distributions corresponding to single power law PDFs of $q_0$, we can recover the commonalities of a realistic orbital period distribution (with the latter a linear combination of the former). The power series should not be dominated by terms with large positive exponents as white dwarf pollutants, which potentially correspond to scattering into the Roche limit ($q_0<r_{\mathit{Roche}}$), is not rare \citep{2003ApJ...596..477Z,2010ApJ...722..725Z,2014A&A...566A..34K,2019MNRAS.487..133W,2023MNRAS.518.3055O,2024MNRAS.527.8687O,2024MNRAS.531L..27M}. The exponents are not necessarily integers as the realistic PDF may already contain a power law(s).

The tidal evolution stage of the planetesimal is quantified by $T_pt$, such that we cannot distinguish a planetesimal with strong tidal response but evolves for a short time from its counterpart with weak tidal response but evolves for a long time. When we vary the tidal evolution time as a free parameter, one can equivalently consider that we vary $T_p$ (or both $T_p$ and $t$ provided that $T_pt$ stays the same).

The effect of each parameter on the probability distribution is plotted in Fig.\ref{CDF plot} and summarized below:

\begin{itemize}
    \item Most importantly, varying the free parameters listed above do not affect the qualitative behaviours of the resultant orbital period distribution: the presence of a peak at the short-period nearly circularized orbit ($\sim 10\,\rm hr$--$1\,\rm day$) and the increasing tail towards long-period highly eccentric orbits ($\sim 100\,\rm day$) on the logarithmic scale. Therefore, these features should persist after a linear combination of individual orbital period distributions corresponding to different power law $q_0$ distributions and different sets of free parameters.
    \item If the planetesimal is scattered earlier and experiences longer tidal evolution/evolves faster under tide, there is a higher probability for short-period (nearly) circularized orbits (upper-left panel).
    \item If the planetesimal leaves the scattering zone at a larger apocentre distance, the probability ratio of being on a (nearly) circularized orbit to its counterpart of being on a partially circularized orbit increases (upper-right panel). 
    \item Increasing the density alone, which reduces the Roche limit, extends the allowed initial pericentre distance towards smaller values where tidal circularization is much faster, hence shifting the orbital period distribution towards shorter orbital period with larger contributions from the (nearly) circularized orbits (lower-left panel).
    \item Including a density/Roche limit distribution of the scattered planetesimal leads to a decaying tail towards the shortest orbital period that is only reachable by the planetesimals with large density/small Roche limit (lower-left panel, black solid line, also see Fig.\ref{integrated distribution plot}).
    \item The initial pericentre distance distribution has non-negligible impact on the orbital period distribution quantitatively: the contributions from the (nearly) circularized orbits to the probability distribution becoming smaller with a decreasing likelihood of being scattered to a smaller pericentre distance; but not qualitatively (lower-right panel, see Appendix E for other forms of distributions).
\end{itemize}

We acknowledge that when tidal circularization is too slow, the cumulative distributions do not necessarily start at the circularization period of a planetesimal scattered to the Roche limit as is the case in Fig.\ref{CDF plot}. Instead, it may truncate (reach 0) at a longer orbital period and the peak at short-period (nearly) circularized orbits does not present (Appendix D). 

\subsection{Tidal disruption or tidal circularization?}\label{roche limit discussion}

\begin{figure}
\includegraphics[width=0.45\textwidth]{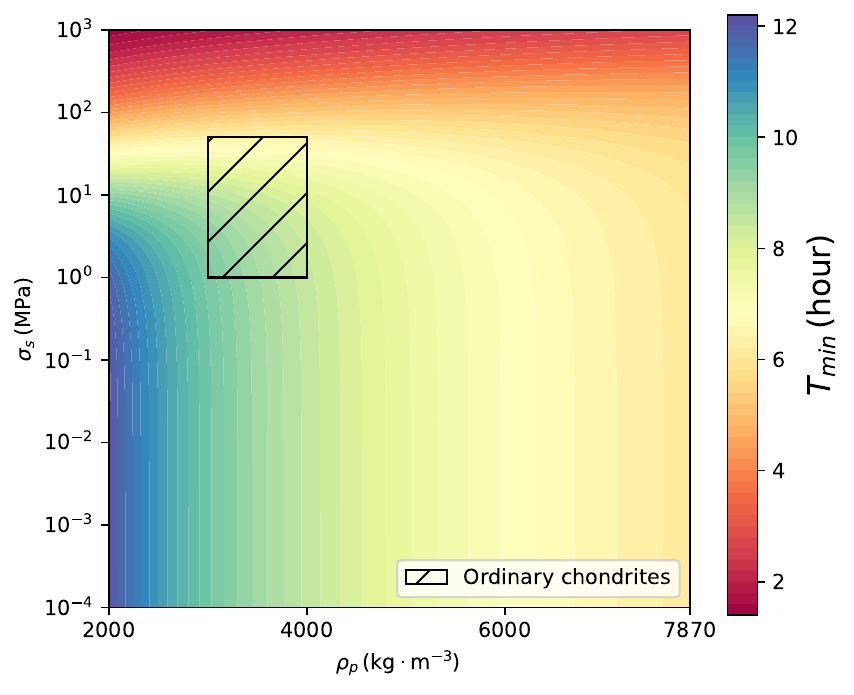}
\caption{The minimum orbital period that a spherical and rigid planetesimal can circularize to without being tidally disrupted in $\rho_p$--$\sigma_s$ space. Other system properties are listed in Table \ref{tidal parameters table}. The x-axes terminate at the density of iron and the hatched rectangle represents the estimated property range of ordinary chondrites (H chondirtes, L chondrites, and LL chondirtes) \citep{2020M&PS...55..962P}.}
\label{roche limit plot}
\end{figure}

Planetesimals scattered sufficiently close to the white dwarf are torn apart by the differential tidal forces. This work focuses on those planetesimals scattered insufficiently close to be torn apart, but sufficiently close for tidal forces to circularize them. 

This work assumes that the Roche limit delineates the minimum pericentre of a planetesimal to escape tidal disruption. As planetesimals scattered to highly eccentric orbits tidally circularize to an semi-major axis around twice their initial pericentre distance ($2q_0$, Eq.\ref{a-e relation from constant L}) and the circularization timescale of the planetesimal follows $\tau_{\mathit{cir}}\propto q_0^{7.8}$ (Section \ref{circularization timescale}), the minimum orbital period a planetesimal can circularize to, and the minimum time required for tidal circularization, are closely related to the lower limit of $q_0$: the Roche limit ($r_{\mathit{Roche}}$). Here we discuss the validity of this assumption and its implications for the model predictions. 

In the previous sections, we estimate the Roche limit assuming that the planetesimal possesses an ultimate tensile strength of 0.1\,MPa (Table \ref{tidal parameters table}), roughly corresponding to the lower limit of the measured meteroites samples \citep{2020M&PS...55..962P}. We now discuss the effect of this parameter choice, together with the uncertainties in the ultimate tensile strength, and propose potential scenarios involving the variations of the $r_{\mathit{Roche}}$ during tidal evolution.

In Fig.\ref{roche limit plot}, we plot the minimum orbital period a planetesimal with $R_p=100\,\rm km$ can circularized to without undergoing tidal disruption ($T_{\mathit{min}}$), a representation of the Roche limit, as a function of planetesimal density ($\rho_p$) and ultimate tensile strength ($\sigma_s$). We stress that $T_{\mathit{min}}$ is insensitive to $Q_0$ (Eq.\ref{roche limit approximated equation}) and $M_*$ ($T_{\mathit{min}}\propto M_*^{-\frac{1}{2}}r_{\mathit{{Roche}}}^{\frac{3}{2}}$ and $r_{\mathit{{Roche}}}\propto M_*^{\frac{1}{3}}$) since $r_{\mathit{Roche}}\ll Q_0$. At $\sigma_s\lesssim 0.1\,\rm Mpa$, the planetesimal is in the self-gravity dominant regime where $T_{\mathit{min}}$ is nearly independent of $\sigma_s$, above which the increase in $\sigma_s$ can significantly reduce $T_{\mathit{min}}$. Fig.\ref{roche limit plot} shows that an orbital period below $\sim 6\,\rm hr$ can hardly be explained by a tidally circularized planetesimal with high density alone: ultimate tensile strength is always required to avoid tidal disruption.

The hatched rectangular regions in Fig.\ref{roche limit plot} corresponds to the measured values of Solar System ordinary chondrites \citep{2020M&PS...55..962P}, which are the most abundant meteorites and may originate from main belt asteroids \citep{2009Icar..200..698N,2014A&A...567L...7N}. A transiting period around/above the predicted $T_{\mathit{min}}$ of ordinary chondrites, $T\gtrsim 6\,\rm hr$, can be explained by a tidally circularized planetesimal with properties lying in the range of main belt asteroids. On the other hand, a transiting period significantly below 6\,hr potentially indicates the existence of other physical processes (e.g., partial tidal disruption, collisions, gravitational instability) and/or a planetesimal with properties distinguishable from our understanding of rocky planetesimals. We acknowledge that the effect of ultimate tensile strength decays with the size of the body (Eq.\ref{roche limit approximated equation}) such that a planetesimal with a smaller size can be circularized to a shorter orbital period without undergoing tidal disruption when other conditions remain identical. Furthermore, if the planetesimal is non-spherical, it undergoes tidal disruption at a larger distance compared to its spherical counterpart (see Appendix G).

\subsubsection{The scale effect of ultimate tensile strength}

The ultimate tensile strength potentially suffers from scale effect such that $\sigma_s$ declines with the size of the body due to the increasing defects/cracks following the relation \citep{2021AcAau.189..465A}:

\begin{equation}
\sigma_s\propto R^{-3\alpha_s},   
\end{equation}

\noindent where $\alpha_s$ ranges from 0.1 to 0.7 \citep{https://doi.org/10.1111/j.1945-5100.2011.01247.x}. Compared to a cm-sized fragment of the planetesimal, there is at least a factor of $\sim 100$ reduction in the ultimate tensile strength for a planetesimal with $R_p=100\,\rm km$. This reduction factor brings the hatched region corresponding to the properties of ordinary chondrites in Fig.\ref{roche limit plot} to self-gravity dominant regime where $T_{\mathit{min}}\gtrsim 8\,\rm hr$.

We acknowledge that the declining trend in ultimate tensile strength with sample size remains uncertain, as the ultimate tensile strength may be more closely related to the composition, mineralogy and thermal history of the body, which vary a lot even within the same group of bodies \citep{2020M&PS...55..962P}, and are clearly distinguished between the experimentally measured meteorite samples and planetesimals we consider in this study. We highlight the importance of this parameter for the results of this study, despite the large uncertainties of the value it would take in an exoplanetary system. 

\subsubsection{Other potential scenarios}

In this study, we assume that the Roche limit corresponds to a hard cut-off, hence neglecting planetesimals scattered inside the Roche limit. In reality, tidal disruption of a planetesimal may take multiple orbits \citep{2011ApJ...732...74G,2017MNRAS.465.1008V}, especially when the planetesimal only enters its Roche limit briefly near the pericentre distance. For a realistic object, tidal disruption may also depend on the density gradient, chemical gradient and the rheology of the body (see \citealp{1992Icar...95...86S} for an example of an object composed of viscous fluid). 

There may exist a scenario where the planetesimal undergoes partial tidal disruption as well as tidal evolution. As the pseudo-synchronous spin rate decays (Fig.\ref{tidal a e weq plot}, upper panel) and the pericentre distance increases with tidal evolution, the planetesimal migrates away from the shrinking Roche limit and tidal disruption ceases. 

Meanwhile, thermal evolution of the planetesimal accompanied with tidal evolution may alter its shape, and the ultimate tensile strength, which potentially correlates with size, porosity, composition, mineralogy and thermal history \citep{2020M&PS...55..962P}, altering the Roche limit. The ultimate tensile strength may be correlated with the degree of deformation of the object (for instance, consider the correlation between ultimate tensile strength and Young's modulus). If $\sigma_s$ is reduced and the planetesimal becomes more deformed (e.g., due to melting) during tidal evolution, the resultant increase in the Roche limit potentially re-triggers tidal disruption. 

\subsection{Comparison to the observed transiting systems}\label{period distribution discussion}

\begin{figure*}
\includegraphics[width=0.8\textwidth]{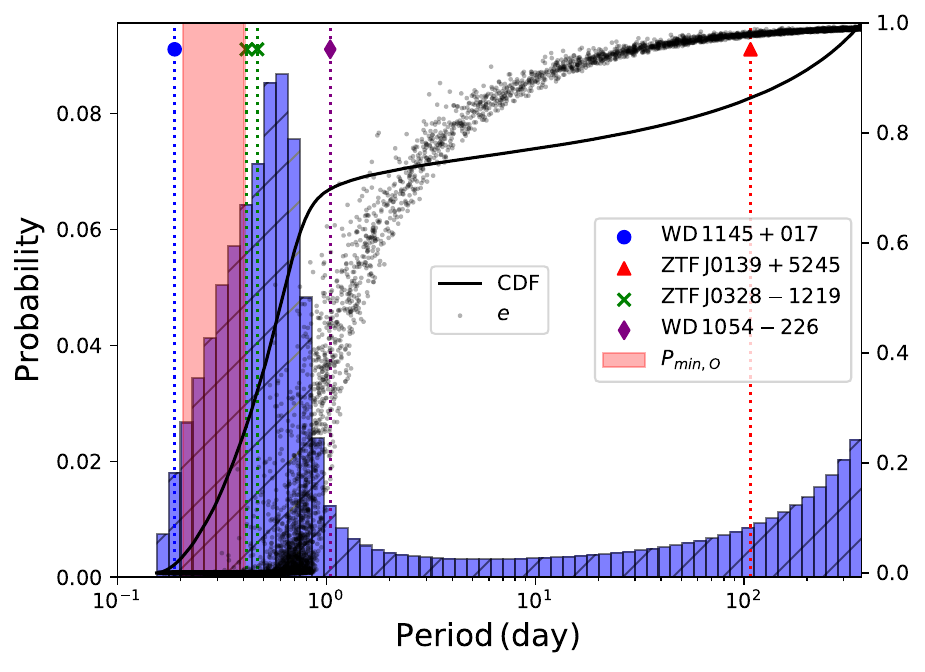}
\caption{The probability distribution of orbital periods of a population of planetesimals, orbiting a population of white dwarfs under tidal evolution. Full details can be found in Section \ref{period distribution discussion}.
Both the x-axis and the bin width are on the logarithmic scale. The black solid line is the cumulative distribution (right axis). The black dots are 10000 randomly sampled orbital eccentricities (right axis). The inferred period seen in the optical transits of WD\,1145+017 (blue circle), ZTF\,J0139+5245 (red triangle), ZTF\,J0328-1219 (green crosses) and WD\,1054–226 (purple diamond) are shown \citep{2015Natur.526..546V,2020ApJ...897..171V,2021ApJ...917...41V,2022MNRAS.511.1647F}. The red shaded region is the range of minimum orbital period achievable by Solar System ordinary chondrites ($P_{\mathit{min,O}}$) as is predicted in Fig.\ref{roche limit plot} (the hatched region).}

\label{integrated distribution plot}
\end{figure*}

\begin{table*}
\begin{center}
\begin{tabular}{|c|c|c|c|c|c| } 
 \hline
System&Mass\,($M_{\odot}$)&Cooling age\,(Myr)& Period&Predicted eccentricity& Predicted normalised probability\,(\%)\\
\hline
WD\,1145+017&0.6&175&4.5\,hour& 0 & 1.8\\
\hline
ZTF\,J0139+5245&0.52&-&107.2\,day&$\sim 0.98$&0.9\\
\hline
ZTF\,J0328-1219&0.73&1840&9.9\,hour&$\sim 0$& 6.4\\
&&&11.2\,hour& $\sim 0$& 7.1\\
\hline
WD\,1054–226 &0.62&1300& 25\,hour & $\sim 0.2$--0.6& 1.2\\
\hline 

\end{tabular}
\end{center}
\caption{A list of the properties of the observed transiting systems \citep{2015Natur.526..546V,2020ApJ...897..171V,2021ApJ...917...41V,2022MNRAS.511.1647F}, together with the predicted orbital eccentricity and normalised (relative) probability (which, qualitative speaking, is insensitive to the number of bins of the histogram) based on the simulated distribution in Fig.\ref{integrated distribution plot}.}
\label{transit table}
\end{table*}

This work presents a model that predicts a population of planetesimals on a wide range of orbital periods around the white dwarf, potentially responsible for the observed transits. In this section, the single-body probability distribution obtained in Section \ref{period distribution} is expanded to a population of planetary systems, accounting for a variety of two-body system properties and tidal evolution stages, thus, mimicking the diversity of the chanced observations of transits. The predicted probability distribution is compared to the current observations in Fig.\ref{integrated distribution plot} and summarized in Table \ref{transit table}.

To deduce the orbital parameters of a planetesimal around a white dwarf under tidal evolution, one need to know 1. the tidal evolution track of the planetesimal in $a$--$e$ space (constrained by initial orbital parameters $q_0$ and $Q_0$) and 2. the position of the planetesimal on the tidal evolution tracks (the tidal evolution stage). Therefore, the corresponding minimum set of distributions required to account for the diversity of transiting systems is: 1. an initial orbital parameter distribution ($q_0$, $Q_0$) and 2. a tidal evolution stage distribution (multiple of tidal evolution time and tidal response, $T_pt$). 

Fig.\ref{integrated distribution plot} shows the orbital period distribution (left axis, histogram), together with a sampled eccentricity (black dots, right axis) and the cumulative distribution (black line, right axis). The normalised probability distribution is calculated by considering a uniform range of potential evolution times (from $t=0$ to $t=500\,\rm Myr$, with an interval of 5\,Myr) and planetesimal densities ($\rho_p=2000$--8000\,$\rm kg\cdot m^{-3}$ with an interval of 500\,$\rm kg\cdot m^{-3}$). We assume that the ultimate tensile strength $\sigma_s$ increases linearly with $\rho_p$ from $1\times 10^5$\,Pa to $4\times 10^{8}$ $\,\rm Pa$ (motivated by \citealp{2019P&SS..165..148O,2020M&PS...55..962P}). The planetesimals are assumed to possess two $Q_0$, 3\,AU and 5\,AU with equal probability. Other parameters are identical to those listed in Table \ref{tidal parameters table}. Although our choices are arbitrary, one may expect that the predictions remain qualitatively similar for a range of free parameter distributions (see Section \ref{free parameters}). 

The observed transiting systems are generally consistent with our predictions: 1. more frequent short-period orbits plus a small enhancement towards longer period and 2. a higher probability towards the transition around the (nearly) circularized and partially circularized orbits. However, current observations are too few and do not form an unbiased distribution to compare with the simulated distribution. We note here that in reality, the probability of detecting a transiting planetesimal with a measured period may differ from the probability of having a planetesimal on such an orbital period (see Appendix F). Here we discuss each system individually.

\subsubsection{WD\,1145+017}

 The transiting planetesimal around WD\,1145+017 is most likely in a circularized orbit, with a normalised probability of $1.8\%$. The 4.5\,hr period is very close to the lower limit of the probability distribution. Within the context of this model, planetesimals are only circularized onto such a short-period orbit, if they undergo partial tidal disruption or possesses high ultimate tensile strength (Section \ref{roche limit discussion}). According to Fig.\ref{roche limit plot} $\sigma_s\gtrsim 130\,\rm MPa$, which is of similar order of magnitude as the measured values for iron meteorites \citep{2020M&PS...55..962P,2021AcAau.189..465A}, is required, if the planetesimal with $R_p=100 \,\rm km$ is to avoid tidal disruption. Alternatively, if the planetesimal possess similar properties to ordinary chondritres, it must be smaller than $\sim 50\,\rm km$ to avoid tidal disruption, well below the predicted value \citep{2016MNRAS.458.3904R}.
 
 Interestingly, if the planetesimal orbiting WD\, 1145+0117 were to have the maximum ultimate tensile strength measured for meteorites, $\sim 400\,\rm MPa$ \citep{2019P&SS..165..148O,2020M&PS...55..962P}, the maximum radius of a planetesimal that would avoid tidal disruption is $\sim 180\,\rm km$, which is similar to the predicted size via the drift periods \citep{2016MNRAS.458.3904R}, potentially suggesting that the planetesimal already underwent partial tidal disruption.
 
We acknowledge that WD\,1145+017 is an extreme of our model, with strict constraints on its properties (such an object may be quite rare, but not impossible, see e.g., \citealp{2019Sci...364...66M}). Whilst this object can be explained in the context of the theory presented here, it does pose the question of why such an extreme case represents the first discovery of a white dwarf with optical transits. One possible explanation is the enhanced transit probability and signal-to-noise ratio towards smaller initial pericentre distance/smaller circularization period (Appendix F). Additionally, the exact tidal response of a body with the properties required to explain WD\,1145+017, is not well understood and we acknowledge that some uncertainties remain in whether such a body can tidally circularize within the cooling age of WD\,1145+017.

\subsubsection {ZTF\,J0139+5245} 

The transiting period of ZTF\,J0139+5245 is best explained by a highly eccentric orbit with $e\sim 0.98$ and lies in the increasing trend of the predicted probability distribution towards long period, with a normalised probability of $0.9\%$.  We postulate four reasons why long-period orbits may dominate this system (based on Section \ref{free parameters}): 1. planetesimals are rarely scattered close to the white dwarf, 2. planetesimals originate from close to the white dwarf (a close-in perturber due to e.g., planet-planet scattering or common envelope event), 3. planetesimals generally evolve slower under tides compared to our predictions, and 4. scattering takes longer (due to e.g., low mass perturbers).

\subsubsection{ZTF\,J0328-1219}

The transiting periods of ZTF\,J0328-1219 which most likely correspond to two (nearly) circularized orbits are close to the peak of the predicted probability distribution, with a normalised probability of $6.4\%$ and $7.1\%$, respectively, corresponding to a combined probability of $\sim 9\%$ of having two such planetesimals at the same time. As the peak of the distribution indicating the transition between (nearly) circularized orbits and partially circularized orbits shifts towards a longer orbital period with increasing $T_pt$ (\ref{period distribution}), the maximum probability of observing the two transiting periods around ZTF\,J0328-1219 is reached for a earlier tidal evolution stage: shorter tidal evolution time/weaker tidal response ($8.8\%$ and $8.5\%$ for $t=0$--100\,Myr).

According to our model, properties of ordinary chonrites is sufficient for planetesimals to circularize onto the 9.9\,hr and 11.2\,hr orbits without undergoing tidal disruption, even when including scale effect and accounting for non-spherical bodies (see Appendix G).

\subsubsection{WD\,1054–226}

The transiting period of WD\,1054–226 most likely corresponds to a partially circularized orbit, whose eccentricity ranges from $\sim 0.2$ to $\sim 0.6$ according to our model, with a probability of $1.3\%$.

We cannot constrain the properties of planetesimals around ZTF\,J0139+5245 and WD\,1054–226 as there is no direct relation between orbital period and initial pericentre distance for partially circularized orbits.

\subsection{Implications}\label{implication}

In this work, we study a potential evolution pathway of white dwarf planetary systems: planetesimals scattered barely outside the Roche limit undergoes tidal circularization (declining semi-major axis and eccentricity). This has crucial consequences for our view of planetary systems orbiting white dwarfs, with more objects potentially populating the inner regions closer to the white dwarf than previously thought. If these planetesimals control the periodic signals found in optical photometric monitoring of many white dwarfs, this scenario has testable observational consequences. Future observations should find a distribution of periods in the systems with optical transits that accumulates at both short-period (nearly) circularized orbits ($\sim 10 \,\rm hr$), and at long-period highly eccentric orbits ($\gtrsim 100\,\rm day$). This work highlights the importance of current and future observing facilities finding new candidate transiting systems, together with follow-up observations confirming their transiting periods. Those objects transiting with short periods that are likely on near-circular orbits has the following advantages:

\begin{itemize}
    \item They provide the best potential tests for models of high eccentricity tidal migration.
    \item Their orbital period may help constrain where the tidally evolved planetesimals originate from and their properties.
    \item Practically, it is easier to constrain their orbital periods using dedicated high-speed photometric follow-up observing campaign in comparison to their long-period counterparts.
\end{itemize}  

Optical transits around white dwarfs are only found in systems where planetary material has been accreted. Given the active dust production potentially required to produce the observed transit signatures, these systems provide an important clue in our understanding of how planetary material is accreted by white dwarfs. A better understanding of the evolution of white dwarf planetary systems will benefit the interpretation of planetary material accreted by white dwarfs that are used to probe the composition of exoplanetary bodies.

\section{Conclusions}\label{conclusion}

Planetary systems around white dwarfs are important targets to investigate the composition of exoplanetary systems. Transits of a handful of polluted white dwarfs provide key observational evidence regarding how planetary material is accreted. A number of mysteries remain regarding the exact details of the accretion process.

In this work, we investigate tidal circularization of planetesimals scattered close to white dwarfs exterior to the Roche limit. Tidally evolved planetesimals are predicted to exist on a wide range of orbital periods, potentially linked to the photometric variability of a handful of polluted white dwarfs, with the orbital periods of the planetesimals controlling the periodic signals seen in the optical data. 

Our simulations predict that under tidal evolution, there exists a population of planetesimals on short-period (nearly) circularized orbits around white dwarfs (peaking at $\sim 10\,\rm hour$--$1\,\rm day$), potentially represented by systems such as WD\,1145+017, ZTF\,J0328-1219 and WD\,1054–226. Alongside these, there is a population of planetesimals on long-period highly eccentric orbits ($\sim 100\,\rm day$) such as seen for ZTF\,J0139+5245 exist, together with a low probability of finding planetesimlas on orbital periods of $\sim 1 \,\rm day$--$100\,\rm day$. 
 
While the orbital periods of most transiting systems can be explained by tidally evolved planetesimals with properties similar to Solar System ordinary chondrites, in order to avoid tidal disruption, the planetesimal on the 4.5\,hour period around WD\,1145+017 must possess an ultimate tensile strength of the same order of magnitude as iron meteorites.

Currently, this field is limited by the small number of transiting systems characterised so far. Whilst the current observations are in-line with the theory presented here, the true test will be comparison of the prediction orbital distribution with the results of the many current and future optical photometric monitoring surveys, including ZTF and Roman Observatories. 

\section*{Acknowledgements}

We would like to thank Dimitri Veras, Siyi Xu and Zach Vanderbosch for valuable discussions regarding various aspects of this paper. AB and LKR acknowledges the support of a Royal Society University Research Fellowship, URF\textbackslash R1\textbackslash 211421. YL acknowledges the support of a STFC studentship. LKR acknowledges support of an ESA Co-Sponsored Research Agreement No. 4000138341/22/NL/GLC/my = Tracing the Geology of Exoplanets


\section*{Data Availability}

Codes and data used in this work are available upon reasonable request to the author, Yuqi Li. 



\bibliographystyle{mnras}
\bibliography{example} 




\appendix


\bsp	
\label{lastpage}
\end{document}


\maketitle
\appendix

\section{Simplified tidal evolution equations}\label{simplified tidal equation appendix}

In order to illustrate the properties of the CTL model \citep{1981A&A....99..126H,2007A&A...462L...5L,2010A&A...516A..64L,2010ApJ...725.1995M,2011A&A...535A..94B,2011A&A...528A..27H,2012ApJ...751..119B,2012ApJ...757....6H,2022ApJ...931...11G,2022ApJ...931...10R,2023ApJ...948...41L} more clearly, we simplify the coupled tidal evolution equations based on the fact that the orbital angular momentum of a white dwarf-planetesimal system is conserved:

\begin{equation}\label{constant L equation appendix}
\begin{aligned}
a(1-e^2)=\frac{2q_0Q_0}{q_0+Q_0},
\end{aligned}    
\end{equation}

\noindent and the fact that the planetesimal reaches pseudo-synchronization and spin-orbit (mis)alignment rapidly. In this case the coupled tidal evolution equations are simplified to: 

\begin{equation}\label{modifed tidal e equation appendix}
\frac{de}{dt}=9GT_pO_{-8}F_e(e),    
\end{equation}

\begin{equation}\label{modifed tidal a equation appendix}
\frac{da}{dt}=2GT_pO_{-7}F_a(e),   
\end{equation}

\begin{equation}\label{equilibrium spin zero obliquity appendix}
\omega_{\mathit{eq}}=\left[G(M_p+M_*)\right]^{\frac{1}{2}}\left(\frac{q_0+Q_0}{2q_0Q_0}\right)^{\frac{3}{2}}\frac{f_2(e)}{f_5(e)}, 
\end{equation}

\begin{equation}
\epsilon_{\mathit{p}}=
    \begin{cases}
    0 & 0\leq \epsilon_{\mathit{p,0}}< \frac{\pi}{2}\\
    \pi & \frac{\pi}{2}<\epsilon_{\mathit{p,0}}\leq \pi
    \end{cases}  
    ,
\end{equation}

\noindent where $T_p$, $O_l$ $F_e(e)$ and $F_a(e)$ are given by:

\begin{equation}
T_p\equiv \frac{K_p(M_p+M_*)M_*R_p^5}{M_p}\approx\frac{K_pM_*^2R_p^5}{M_p}\propto \frac{K_pM_*^2R_p^2}{\rho_p},
\end{equation}

 \begin{equation}
O_l\equiv \left(\frac{2q_0Q_0}{q_0+Q_0}\right)^{l},     
\end{equation}

\begin{equation}\label{Fe equation}
\begin{aligned}
F_e(e)=e(1-e^2)^{\frac{3}{2}}\left[\frac{11}{18}\frac{f_2(e)f_4(e)}{f_5(e)}-f_3(e)\right],   
\end{aligned}   
\end{equation}

\begin{equation}\label{Fa equation}
\begin{aligned}
F_a(e)=(1-e^2)^{-\frac{1}{2}}\left[\frac{f_2^2(e)}{f_5(e)}-f_1(e)\right].   
\end{aligned}   
\end{equation}

We can deduce the following properties of the CTL model:

\begin{itemize}
    \item Tidal evolution track in $a$--$e$ space is determined by $F_e(e)$/$F_a(e)$, which means that it is fixed for a given set of $(q_0,Q_0)$.
    \item Tidal evolution rates in both $a$ and $e$ are scaled by $T_p$, such that for a fixed $(q_0,Q_0)$, $T_pt$ determines the tidal evolution stage (tidal evolution is simultaneous in $T_pt$ space).
    \item At identical $e$ and for $q_0\ll Q_0$, $O_l$ indicates that a planetesimal starts at a smaller $q_0$ evolves much faster.
    \item $\frac{de}{dt}\propto F_e(e)\leq 0$ has its minimum at $e\approx 0.658$, converging to 0 at both $e\rightarrow 0$ and $e\rightarrow 1$,
    \item $\frac{da}{dt}\propto F_a(e) \leq 0$ decreases monotonically with $e$, converging to 0 at $e\rightarrow 0$.
    \item $\frac{dT}{dt}\propto a^{\frac{1}{2}}\frac{da}{dt}\propto (1-e^2)^{-\frac{1}{2}}F_a(e)\leq 0$ has the same qualitative behaviour as $F_a(e)$ while decreasing more strongly with $e$.
    \item Pseudo-synchronous spin rate of the planetesimal decreases along the tidal evolution track.
\end{itemize}

\section{Analytical tidal circularization timescale}\label{tidal circularization timescale appendix}

One can approximate the circularization timescale by analyzing $|\frac{a}{\dot{a}}|_0$ in the limit of $q_0\ll Q_0$ ($e_0=\frac{2Q_0}{q_0+Q_0}-1\rightarrow 1$). We expand the terms in the square brackets in Eq.\ref{Fa equation} at $e_0=1$ such that:

\begin{equation}
\left[\frac{f_2^2(e_0)}{f_5(e_0)}-f_1(e_0)\right]\approx -\frac{4059}{320}-\frac{264(e_0-1)}{5}. 
\end{equation}

We then expand at $\frac{q_0}{Q_0}=0$ and get:

\begin{equation}\label{circularization timescale}
\begin{aligned}
\left|\frac{\dot{a}}{a}\right|_0&\approx {2GT_p}\left(\frac{2q_0Q_0}{q_0+Q_0}\right)^{-8}(1-e_0^2)^{\frac{1}{2}}\left[\frac{4059}{320}-\frac{264(1-e_0)}{5}\right]\\& \approx {2GT_p}\frac{4059}{320}\left(2q_0\right)^{-8}\left(4\frac{q_0}{Q_0}\right)^{\frac{1}{2}}\\
&\approx {0.2GT_p}q_0^{-7.5}Q_0^{-0.5},
\end{aligned}
\end{equation}

\noindent such that $\tau_{\mathit{cir}}\approx \frac{5}{GT_p} q_0^{7.5} Q_0^{0.5}$, increasing much more rapidly with $q_0$ than $Q_0$ and $T_p$.

\section{Interpretations of orbital period distribution}\label{analytical orbital period distribution}

In this section we focus on the case where $e\rightarrow 0$ and $q_0\ll Q_0$. We will use the fact that on the logarithmic scale, the widths of the bins of the orbital period distribution histogram $\Delta T$ satisfies $\frac{\Delta T}{T}=\mathrm{constant}$, while on the linear scale, $\Delta T$ is a constant.

Eq.\ref{constant L equation appendix} implies:

\begin{equation}\label{delta q equation}
 2\Delta q_0=(a+\Delta a)[1-(e+\Delta e)^2]-a(1-e^2),
\end{equation}

\noindent where $\Delta a=a(q_0+\Delta q_0)-a(q_0)$, and $\Delta e=e(q_0+\Delta q_0)-e(q_0)$.

Kepler's third law implies that:

\begin{equation}\label{delta a delta T relation}
\begin{aligned}
&a=A T^{\frac{2}{3}},\\
&\frac{\Delta a}{a}=(1+\frac{\Delta T}{T})^{\frac{2}{3}}-1,  
\end{aligned} 
\end{equation}

\noindent where $A$ is a positive proportionality constant.

Eq.\ref{delta q equation} and Eq.\ref{delta a delta T relation} combine to give: 

\begin{equation}\label{period distribution eccentric}
\begin{aligned}
&2\Delta q_0=a(1+\frac{\Delta T}{T})^{\frac{2}{3}}[1-(e+\Delta e)^2]-a(1-e^2),
\end{aligned}   
\end{equation}

\noindent where $\Delta q_0=q_0(T+\Delta T)-q_0(T)$ (for the orbital period distribution histogram, $\Delta T$ is the bin width and $T$ is its inner bin edge). By substituting $a(1-e^2)=2q_0$, Eq.\ref{period distribution eccentric} can be expressed as:

\begin{equation}\label{q T relation}
\begin{aligned}
\frac{\Delta q_0}{q_0}=(1+\frac{\Delta T}{T})^{\frac{2}{3}}\left[\frac{1-(e+\Delta e)^2}{1-e^2}\right]-1.  
\end{aligned}
\end{equation}

After applying the PDF $P(q_0)\propto q_0^{\alpha}$, one can write (for $\alpha \neq -1$):

\begin{equation}\label{P equation}
\begin{aligned}
P([T,T+\Delta T])&=P([q_0,q_0+\Delta q_0])\\&=\int_{q_0}^{q_0+\Delta q_0}P(q_0')dq_0'\\&=N_{\alpha}\left[(q_0+\Delta q_0)^{\alpha+1}-q_0^{\alpha+1}\right]\\& =N_{\alpha} q_0^{\alpha+1}\left|\left(1+\frac{\Delta T}{T}\right)^{\frac{2}{3}\alpha+\frac{2}{3}}\left[\frac{1-(e+\Delta e)^2}{1-e^2}\right]^{\alpha+1}-1\right|\\& =N_{\alpha}\left(\frac{A}{2}\right)^{\alpha+1}T^{\frac{2}{3}\alpha+\frac{2}{3}}\times \\ &\left|\left(1+\frac{\Delta T}{T}\right)^{\frac{2}{3}\alpha+\frac{2}{3}}\left[{1-(e+\Delta e)^2}\right]^{\alpha+1}-(1-e^2)^{\alpha+1}\right|,
\end{aligned}  
\end{equation}

\noindent where $N_{\alpha}$ is a positive normalisation constant for $\alpha$. When $\alpha>-1$, the absolute value signs in Eq.\ref{P equation} can be omitted. A positive $\alpha$ is usually expected as it is less likely for a planetesimal to be scattered to a smaller pericentre distance \citep{2024MNRAS.527.11664} but we will discuss the case where $\alpha<0$ for completeness. 

For $\alpha=-1$, we have:

\begin{equation}
\begin{aligned}
P([T,T+\Delta T])&=P([q_0,q_0+\Delta q_0])\\&=N_{-1} \left[\ln{(q_0+\Delta q_0)}-\ln{q_0}\right] \\& =N_{-1}  \ln\left(1+\frac{\Delta q_0}{q_0}\right)\\& =N_{-1} \ln\left\{\left(1+\frac{\Delta T}{T}\right)^{\frac{2}{3}}\left[\frac{1-(e+\Delta e)^2}{1-e^2}\right]\right\}.
\end{aligned}  
\end{equation}

For (nearly) circularized orbits ($e\rightarrow 0$, $\Delta e\rightarrow 0$, $a=2q_0$), we have, for $\alpha\neq-1$:

\begin{equation}
\begin{aligned}
P([T,T+\Delta T]|e\rightarrow 0)=N_{\alpha}\left(\frac{A}{2}\right)^{\alpha+1} T^{\frac{2}{3}\alpha+\frac{2}{3}}\left|\left(1+\frac{\Delta T}{T}\right)^{\frac{2}{3}\alpha+\frac{2}{3}}-1\right|,
\end{aligned}    
\end{equation}

\noindent where a special point is $\alpha=\frac{1}{2}$ (the fiducial value used in the main text), such that:

\begin{equation}\label{alpha 1/2 cir linear}
\begin{aligned}
P([T,T+\Delta T]|e\rightarrow 0)=N_{\frac{1}{2}}\left(\frac{A}{2}\right)^{\frac{3}{2}} \Delta T,
\end{aligned}    
\end{equation}

\noindent which returns a uniform distribution on the linear scale where $\Delta T=\mathrm{constant}$ (see the zoom-in plot of Fig.6 in the main text). Meanwhile, on the logarithmic scale where $\frac{\Delta T}{T}=\mathrm{constant}$, we have:

\begin{equation}\label{alpha 1/2 cir log}
\begin{aligned}
P([T,T+\Delta T ]|e\rightarrow 0)=B T^{\frac{2}{3}\alpha+\frac{2}{3}},
\end{aligned}    
\end{equation}

\noindent where $B=N_{\alpha}\left(\frac{A}{2}\right)^{\alpha+1}\left|\left(1+\frac{\Delta T}{T}\right)^{\frac{2}{3}\alpha+\frac{2}{3}}-1\right|$ is a constant. For $\alpha=\frac{1}{2}$, the probability distribution is linearly proportional to $T$ for (nearly) circularized orbits, consistent with Fig.6 in the main text. 

For $\alpha=-1$, we have:

\begin{equation}\label{alpha -1 cir}
\begin{aligned}
P([T,T+\Delta T]|e\rightarrow 0)=\frac{2}{3}N_{-1} \ln\left(1+\frac{\Delta T}{T}\right),
\end{aligned}  
\end{equation}

\noindent which returns a uniform distribution on the logarithmic scale where $\frac{\Delta T}{T}=\mathrm{constant}$ (see the lower-right panel of Fig.8 in the main text where the CDF is linear on the logarithmic scale).

The analysis above indicates that the probability distribution of the period for (nearly) circularized orbits help probe the initial pericentre distance distribution. This is expected as $a$ (and hence $T$) is directly related to $q_0$ for (nearly) circularized orbits.

On the other hand, the probability distribution for non-zero eccentricity is not straightforward. We provide a conceptual interpretation here. Imagine two planetesimals starting at identical $Q_0$ but different $q_0$ with $q_0\ll Q_0$, such that they possess similar initial semi-major axis and initial orbital period. Then, when we start the tidal evolution, Eq.\ref{circularization timescale} implies that the $a$ and $T$ of the planetesimal starts at a smaller $q_0$ decays much faster. Therefore, the difference in $T$ of these two planetesimals increase, such that these two planetesimals become more loosely packed in period space. Now, consider a second scenario where the planetesimal starting at a smaller $q_0$ is nearly circularized, such that its $|\frac{dT}{dt}|$ converges to 0, while its counterpart is still on a partially circularized orbit. In this scenario, as $T$ of the planetesimal with a larger $q_0$ continues to decrease towards its minimum (which is larger than its counterpart), the difference in $T$ of these two planetesimals declines and these two planetesimals become more closely packed in period space. The decay of the difference in $T$ continues until the planetesimal starting at a larger $q_0$ is (nearly) circularized, where these two planetesimals are in the most closely packed state (see the lower panel of Fig.4 in the main text). To sum up, tidal circularization looses the packing of orbits in period space at an early tidal evolution stage and then tightens the packing towards short-period (nearly) circularized orbits. Hence, planetesimals tend to pile up towards short-period (nearly) circularized orbits and long-period highly eccentric orbits. These scenarios are shown in Fig.4 and Fig.7 in the main text. In Fig.4, the orbital period difference of two planetesimals starting at different $q_0$ initially increases and then decays during tidal evolution, reaching a constant after both planetesimals are circularized. In Fig.7, the initially closely packed orbits in period space at $t=0$ looses between $T\sim 1\,\rm day$ and $T\sim 100\,\rm day$. Towards $T\lesssim 1 \,\rm day$, the nearly circularized orbits become more closely packed where a given $\Delta T$ covers a larger $\Delta q_0$. These features are consistent with the obtained orbital period distribution in Fig.6 in the main text.

\section{Orbital period distribution at an early tidal evolution phase}\label{short time appendix}

\begin{figure}
\begin{center}
\includegraphics[width=0.9\textwidth]{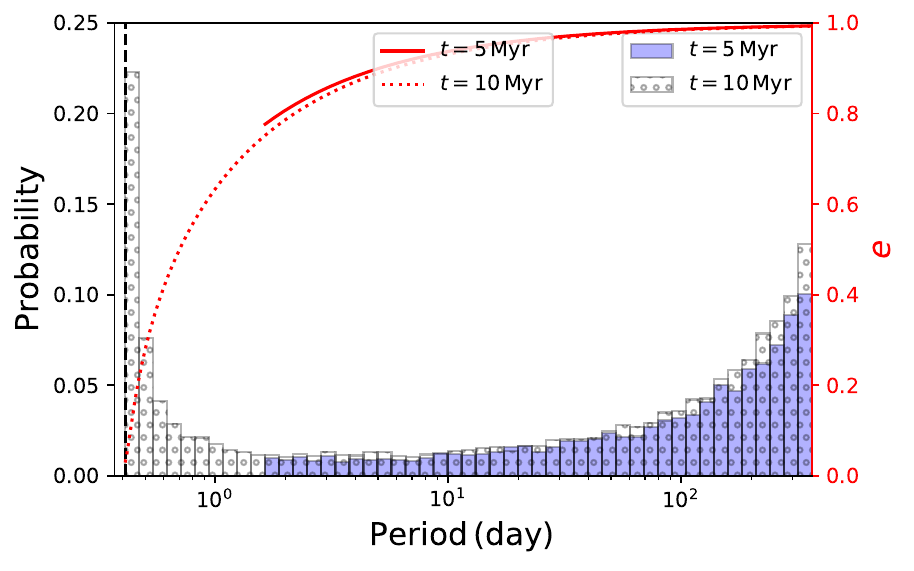}
\caption{Same plot as Fig.6 in the main text, but for shorter tidal evolution timescales, $t=5\,\rm Myr$ and $t=10\,\rm Myr$. The distribution is normalised with respect to $t=5\,\rm Myr$.}
\label{sample_t_subplot_short}    
\end{center}
\end{figure}

\begin{figure}
\begin{center}
\includegraphics[width=0.9\textwidth]{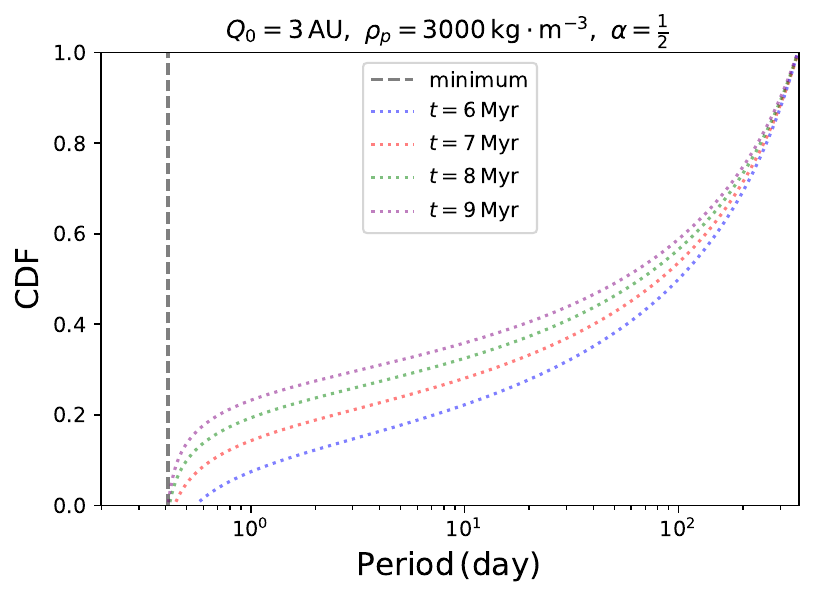}
\caption{Same plot as the upper-left panel of Fig.9 in the main text, but for shorter tidal evolution timescales.}
\label{CDF plot short}    
\end{center}
\end{figure}

In Fig.\ref{sample_t_subplot_short} and Fig.\ref{CDF plot short}, we plot the short tidal evolution time/weaker tidal response (note that $t$ and $T_p$ are degenerate such that tidal evolution is simultaneous in $T_pt$ space) counterparts to Fig.6 and the upper-left panel of Fig.8 in the main text. In Fig.\ref{sample_t_subplot_short}, one can see that the peak towards short period is not present at a tidal evolution time of 5\,Myr, where the minimum eccentricity is around 0.8. At a tidal evolution time of 10\,Myr, the peak appears as the minimum eccentricity approaches 0. Similarly, in Fig.\ref{CDF plot short}, the CDF truncates before the minimum orbital period (dashed line) for a tidal evolution time of 6\,Myr and 7\,Myr. However, at a tidal evolution time of $7\,\rm Myr$, there already exists a small peak (see the gradient of the CDF) towards the shortest orbital period.

\section{Cumulative distribution of orbital period for different forms of initial pericentre distribution}\label{distribution appendix}

\begin{figure}
\begin{center}
\includegraphics[width=0.9\textwidth]{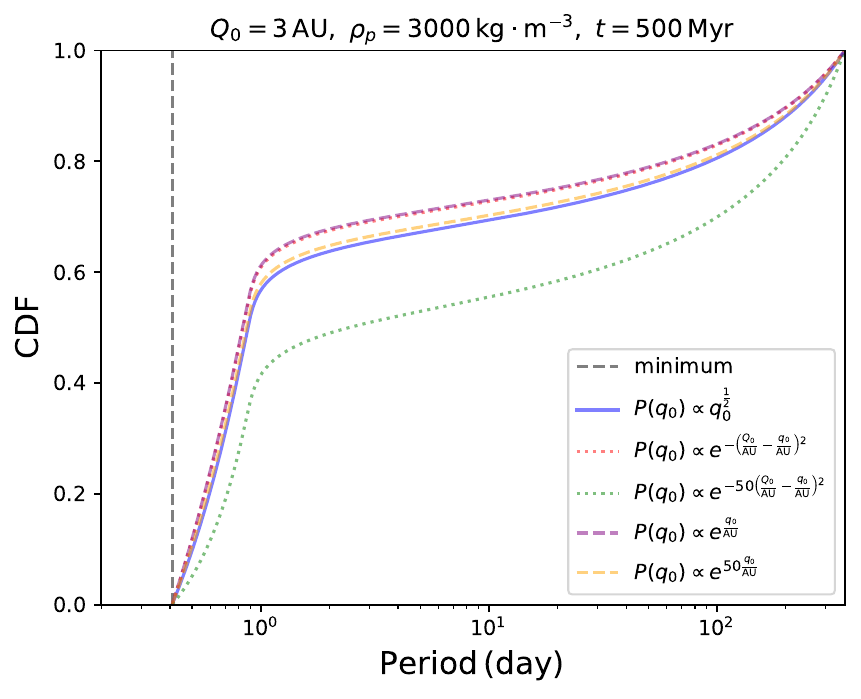}
\caption{The cumulative distribution of planetesimal's period within 1\,year for different forms of initial pericentre distribution.}
\label{CDF plot dis}    
\end{center}
\end{figure}

In Fig.\ref{CDF plot dis}, we plot the resultant orbital period distribution for different forms of probability density function of $q_0$. The qualitative features: a peak at the short-period nearly circularized orbits ($\sim 10\,\rm hr$--$1\,\rm day$) and the increasing tail towards long-period highly eccentric orbits ($\sim 100\,\rm day$) on the logarithmic scale persists.

\section{Observational probability}\label{transit and detection probability}

The transit probability of a planetesimal can be expressed as \citep{2007PASP..119..986B}:

\begin{equation}
P_{\mathit{transit}}=\frac{R_*+R_{\mathit{eff}}}{a(1-e^2)},
\end{equation}

\noindent where $R_{\mathit{eff}}$ is the maximum extent of the planetesimal in the direction perpendicular to the orbital plane that is optically thick. By using Eq.\ref{constant L equation appendix}, one can write:

\begin{equation}
P_{\mathit{transit}}=\frac{(R_*+R_{\mathit{eff}})(q_0+Q_0)}{2q_0Q_0},
\end{equation}

\noindent which decreases with the increase in $q_0$.

The transit duration of a planetesimal can be expressed as \citep{2005ApJ...627.1011T,2008ApJ...679.1566B}:

\begin{equation}\label{transit duration equation}
\tau_{\mathit{transit}}=\frac{2(R_*+R_{\mathit{eff}})r}{\sqrt{G(M_*+M_p)a(1-e^2)}}\sqrt{1-\frac{r^2\cos^2i}{(R_*+R_{\mathit{eff}})^2}},   
\end{equation}

\noindent where $R_{\mathit{eff}}$ depends on the shape, extent of the optically thick region and the orbital inclination $i$, $r=\frac{a(1-e^2)}{1+e\cos{f}}$ is the distance of the planetesimal from the white dwarf with $f$ the true anomaly. 

For a random viewing angle, it is more natural to compute the spatial (uniformly distributed true anomaly) or temporal average (a $f(t)$ distribution with $t$ uniformly distributed between 0 and orbital period) of Eq.\ref{transit duration equation}. In the limit of $i \rightarrow 90^{\circ}$, where $R_{\mathit{eff}}$ represents the maximum extent of the planetesimal in the orbital direction that is optically thick, the averaged Eq.\ref{transit duration equation} is:

\begin{equation}\label{spatial average transit duration}
\begin{aligned}
<\tau_{\mathit{transit}}>_{\mathit{spatial}}&=\frac{2(R_*+R_{\mathit{eff}})}{\sqrt{G(M_*+M_p)a(1-e^2)}}\frac{1}{2\pi}\int_{0}^{2\pi}rdf\\&=\frac{2(R_*+R_{\mathit{eff}})a\sqrt{1-e^2}}{\sqrt{G(M_*+M_p)a(1-e^2)}}\\&=\frac{2(R_*+R_{\mathit{eff}})a^{\frac{1}{2}}}{\sqrt{G(M_*+M_p)}}\propto a^{\frac{1}{2}},
\end{aligned}
\end{equation}

\begin{equation}\label{time average transit duration}
\begin{aligned}
<\tau_{\mathit{transit}}>_{\mathit{time}}&=\frac{2(R_*+R_{\mathit{eff}})}{\sqrt{G(M_*+M_p)a(1-e^2)}}\frac{1}{T}\int_{0}^{T}rdt\\&=\frac{2(R_*+R_{\mathit{eff}})a(1+\frac{e^2}{2})}{\sqrt{G(M_*+M_p)a(1-e^2)}}\\&\propto a^{\frac{1}{2}}(1+\frac{e^2}{2})(1-e^2)^{-\frac{1}{2}}. 
\end{aligned}
\end{equation}

The transit duration is closely correlated with the signal-to-noise ratio \citep{2009ApJ...702..779V}:

\begin{equation}
\mathrm{SNR}=\sqrt{\frac{(d_{\mathit{transit}}N_{\mathit{transit}})^2}{\sum_i \left[N_i^2(\frac{\sigma_w^2}{N_i}+\sigma_r^2)\right]}},   
\end{equation}

\noindent where $d_{\mathit{transit}}$ is the transit depth, $N_{\mathit{transit}}$ is the total number of data points collected during all the transits, $N_i$ is the number of data points collected during the $i$th transit, such that $\sum_iN_i=N_{\mathit{transit}}$, with $i$ ranging from 1 to the nearest integer of $\frac{\tau_{\mathit{obs}}}{T}$, $\sigma_w$ and $\sigma_r$ represents white noise and red noise. Assuming identical observational conditions, we have $N_{\mathit{transit}}\propto \frac{\tau_{\mathit{transit}}}{T}$ and $N_i \propto \tau_{\mathit{transit}}$, $\sum_iN_i^2\propto \frac{\tau_{\mathit{transit}}^2}{T}$. 

We investigate the SNR in two limiting cases. In the white noise dominant limit:

\begin{equation}
 \mathrm{SNR}_{w}\propto \sqrt{N_{\mathit{transit}}}\propto \sqrt{\frac{\tau_{\mathit{transit}}}{T}}\propto \sqrt{\tau_{\mathit{transit}}}a^{-\frac{3}{4}},   
\end{equation}

\noindent which can be expressed in $a$ and $e$ by substituting Eq.\ref{spatial average transit duration} or Eq.\ref{time average transit duration}. 

In terms of the spatial average, the SNR in the white noise dominant limit is:

\begin{equation}
\mathrm{SNR}_{\mathit{w,s}}\propto a^{-\frac{1}{2}}.    
\end{equation}

In terms of the time average, the SNR can be expressed as:

\begin{equation}
\begin{aligned}
\mathrm{SNR}_{\mathit{w,t}}&\propto a^{-\frac{1}{2}}(1+\frac{e^2}{2})^{\frac{1}{2}}(1-e^2)^{-\frac{1}{4}}\\&\propto \left(\frac{q_0+Q_0}{2q_0Q_0}\right)^{\frac{1}{2}}\left(-\frac{e^6}{4}-\frac{3e^4}{4}+1\right)^{\frac{1}{4}},      
\end{aligned} 
\end{equation}

\noindent which increases with the decreasing initial pericentre distance and decreasing orbital eccentricity.

In the red noise dominant limit, the SNR is independent of the transit duration:

\begin{equation}
\mathrm{SNR}_{r}\propto \sqrt{\frac{N_{\mathit{transit}}^2}{\sum_i N_i^2}}\propto \sqrt{\frac{1}{T}}\propto a^{-\frac{3}{4}},
\end{equation} 

\noindent which decreases with the increasing orbital period.

In summary, as short-period (nearly) circularized planetesimals usually start with small $q_0$ (tidal circularization timescale increases most rapidly with $q_0$), they tend to possess large transit probability and large SNR. The SNR generally decays towards an earlier stage of tidal evolution (larger $a$, $T$ and $e$). On the other hand, as a population of planetesimals could be in a wide range of tidal evolution stages due to different tidal evolution time and tidal response, a planetesimal on a long-period highly eccentric orbit (early tidal evolution stage) does not necessarily have a low transit probability.

\section{The Roche limit of non-spherical objects}\label{roche limit appendix}

\begin{figure*}
\begin{center}
\includegraphics[width=0.95\textwidth]{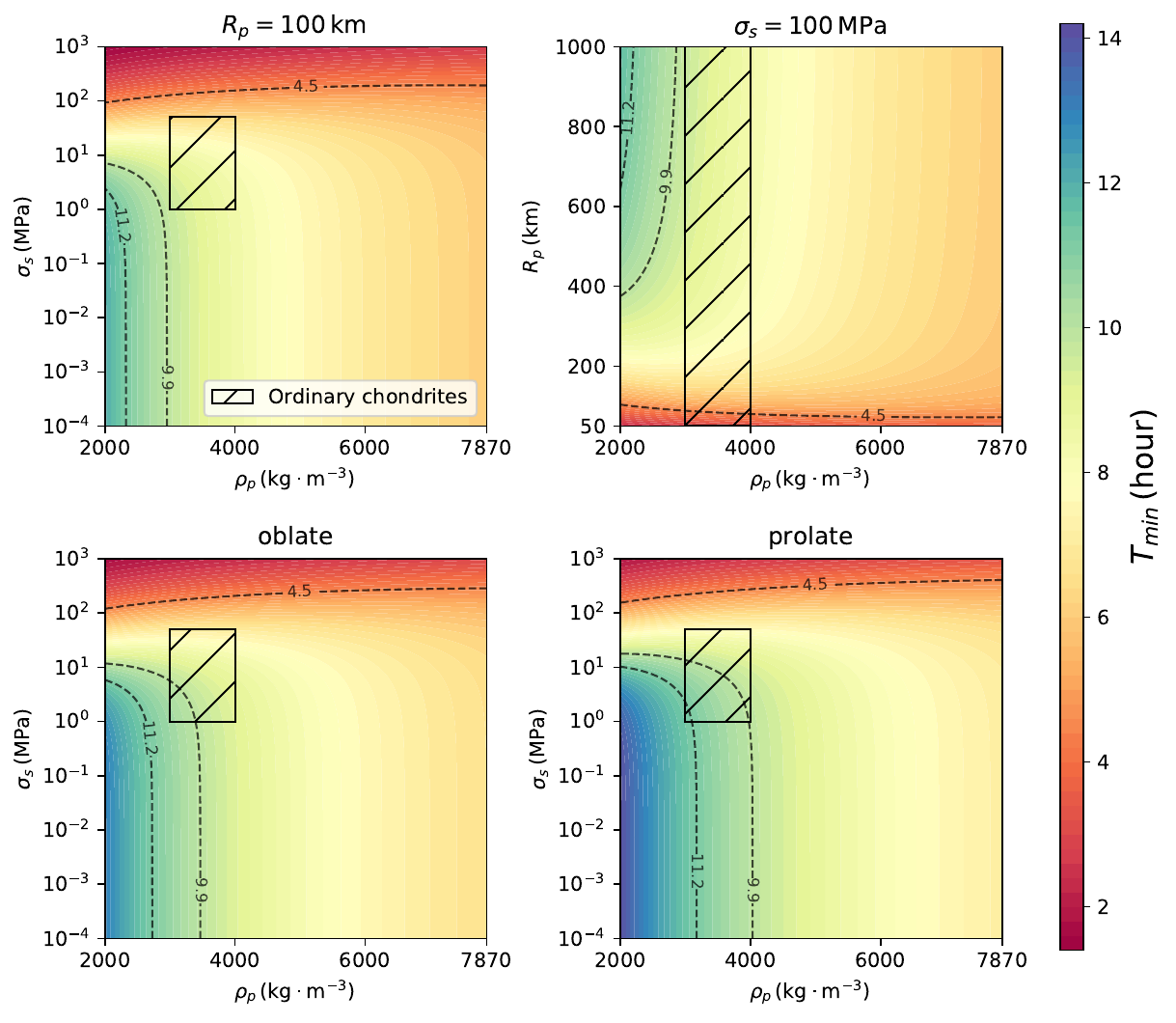}
\caption{The minimum orbital period that a spherical planetesimal can circularize to without being tidally disrupted in $\rho_p$--$\sigma_s$ space (upper-left panel) and $\rho_p$--$R_p$ space (upper-right panel), together with its counterpart of an oblate spheroid (lower-left panel) and prolate spheroid (lower-right panel). Both spheroids have semi-minor axis to semi-major axis ratio $f=0.7$. Other system properties are listed in Table 3 in the main text. The x-axes terminate at the density of iron the hatched rectangle represents the estimated property range of ordinary chondrites (H chondirtes, L chondrites, and LL chondirtes) \citep{2020M&PS...55..962P}. The dashed lines correspond to the inferred periods seen in the optical transits of WD\,1145+017 (4.5\,hr) and ZTF\,J0328-1219 (11.2\,hr and 9.9\,hr) \citep{2015Natur.526..546V,2021ApJ...917...41V}.}
\label{roche limit plot}
\end{center}
\end{figure*}

We consider two simple non-spherical bodies: an oblate spheroid and a prolate spheroid with semi-minor axis (short axis) to semi-major axis (long axis) ratio $f$. We further assume that the semi-minor axis of the non-spherical planetesimal is the spin axis, and that the mass and density of the non-spherical planetesimal is identical to its spherical counterpart. 

Similar to the spherical case (Section 2.2 in the main text), the balance for an oblate spheroid can be expressed as (using the gravitational potential in \citealp{1999Icar..142..525D}):

\begin{equation}
\frac{2GM_*a}{r_{\mathit{Roche}}^3}+\omega_p^2a=GA_o(f)\rho_p a+\frac{\sigma_s \pi fa^2}{M_p},    
\end{equation}

\noindent where $a$ is the semi-major axis of the planetesimal, $M_p=\frac{4}{3}\pi R_p^3 \rho_p=\frac{4}{3}\pi f a^3 \rho_p$, $A_o(f)$ is of the form:

\begin{equation}
A_o(f)=\frac{2\pi f}{(1-f^2)^{\frac{3}{2}}}\tan^{-1}\sqrt{\frac{1}{f^2}-1}-\frac{2\pi f^2}{1-f^2}.
\end{equation}

The approximated Roche limit is hence:

\begin{equation}\label{roche limit oblate equation}
\begin{aligned}
r_{\mathit{Roche, oblate}}=\left[\frac{3.36125GM_*}{GA_o(f)\rho_p+\frac{3\sigma_sf^{\frac{2}{3}}}{4R_p^2\rho_p}}\right]^{\frac{1}{3}},
\end{aligned}
\end{equation}

The balance for an prolate spheroid can be approximated as:

\begin{equation}
\frac{2GM_*a}{r_{\mathit{Roche}}^3}+\omega_p^2a=GA_p(f)\rho_p a+\frac{\sigma_s \pi f^2a^2}{M_p},    
\end{equation}

\noindent where $M_p=\frac{4}{3}\pi R_p^3 \rho_p=\frac{4}{3}\pi f^{2} a^3 \rho_p$, $A_p(f)$ is of the form:

\begin{equation}
A_p(f)=\frac{2\pi f^2}{(1-f^2)^{\frac{3}{2}}}\ln{\frac{1+\sqrt{1-f^2}}{1-\sqrt{1-f^2}}}-\frac{4\pi f^2}{1-f^2},
\end{equation}

The corresponding Roche limit can be approximated as:

\begin{equation}\label{roche limit prolate equation}
\begin{aligned}
r_{\mathit{Roche, prolate}}\approx\left[\frac{3.36125GM_*}{GA_p(f)\rho_p+\frac{3\sigma_sf^{\frac{4}{3}}}{4R_p^2\rho_p}}\right]^{\frac{1}{3}},
\end{aligned}
\end{equation}

Alternatively, the Roche limit can be solved in the same way as Eq.11 in the main text. 

For completeness, we also include the case where the semi-major axis of the prolate planetesimal is the spin axis; 

\begin{equation}
\frac{2GM_*fa}{r_{\mathit{Roche}}^3}+\omega_p^2fa=GA_{p}'(f)\rho_p fa+\frac{\sigma_s \pi fa^2}{M_p},    
\end{equation}

\noindent where $A_{p}'(f)=2\pi-\frac{1}{2}A_p(f)$ and hence:

\begin{equation}
\begin{aligned}
r_{\mathit{Roche, prolate}}\approx\left[\frac{3.36125GM_*}{GA_p'(f)\rho_p+\frac{3\sigma_sf^{-\frac{2}{3}}}{4R_p^2\rho_p}}\right]^{\frac{1}{3}}.
\end{aligned}
\end{equation}

In Fig.\ref{roche limit plot}, we plot $T_{\mathit{min}}$ reached at circularization. The upper-left panel is identical to Fig.10 in the main text. In the upper-right panel of Fig.\ref{roche limit plot}, we plot $T_{\mathit{min}}$ in $\rho_p$--$R_p$ space at $\sigma_{s}= 100\,\rm MPa$. The effect of $\sigma_s$ decays with the increase of $R_p$. Therefore, there may exist a scenario where the planetesimal undergoes partial tidal disruption and size contraction until the ultimate tensile strength is sufficient to support the body. In this case, the final size of the planetesimal on the ciruclarized orbit help constrain the ultimate tensile strength of the body. 

In the lower panels of Fig.\ref{roche limit plot}, we plot the analogues of the upper-left panel for an oblate spheroid (lower-left panel) and a prolate spheroid (lower-right panel) with semi-minor axis (short axis) to semi-major axis (long axis) ratio 0.7, and with the same density and mass as its spherical counterpart in the upper-left panel. We further assume that the semi-minor axis is the spin axis. When other conditions remain the same, a non-spherical planetesimal undergoes tidal disruption at a larger distance compared to its spherical counterpart, with the prolate planetesimal being the weakest against tidal disruption.

\bibliography{appendix}